\documentstyle[12pt]{article}
\def\beq{\begin{equation}} 
\def\eeq{\end{equation}} 
\def\bea{\begin{eqnarray}} 
\def\eea{\end{eqnarray}} 
\def\bq{\begin{quote}} 
\def\eq{\end{quote}}
\def\ra{\rightarrow}

\def\bq{\begin{quote}} 
\def\eq{\end{quote}}
\def\ra{\rightarrow}

\begin{document}

\baselineskip 18pt

\newcommand{\sheptitle}
{Yukawa Textures in String Unified  Models with $SU(4)\otimes O(4)$ Symmetry.}

\newcommand{\shepauthor}
{B. C. Allanach$^1$, S. F. King$^2$, G. K. Leontaris$^{3}$ and S. Lola$^4$}

\newcommand{\shepaddress}
{$^1$Rutherford Appleton Laboratory, Chilton, Didcot, OX11 0QX, U.K.\\ 
$^2$Department of Physics and Astronomy, 
University of Southampton\\Southampton, SO9 5NH, U.K.\\
$^3$Physics Department, University of Ioannina, PO Box 1186,
GR-45110 Ioannina, Greece \\
$^4$Theory Division, CERN, 1211 Geneva 23, Switzerland } 
\newcommand{\shepabstract}
{We discuss the origin of Yukawa textures in the string-inspired and
string derived  models based on the gauge group $SU(4)\otimes
SU(2)_L\otimes SU(2)_R$ supplemented by a $U(1)_X$ gauged family symmetry. 
The gauge symmetries are broken down to those of the minimal supersymmetric
standard model which is the  effective theory below $10^{16}$ GeV.  The
combination of the $U(1)_X$ family symmetry and  the Pati-Salam gauge group
leads to a successful and predictive set of Yukawa textures involving two
kinds of texture zeroes: {\em horizontal} \/and {\em vertical} \/texture
zeroes.
We discuss both symmetric and non-symmetric textures in models of this kind,
and in the  second case perform a detailed numerical fit to the  charged
fermion mass and mixing data. Two of the Yukawa textures allow a low energy
fit to the data with a total $\chi^2$ of 0.39 and 1.02 respectively, for
three degrees of freedom. We also make a first attempt at deriving the non
- renormalisable operators required for the Yukawa textures from string
theory.}

\begin{titlepage}
\begin{flushright}
hep-ph/9610517\\
CERN-TH-96-300\\ IOA-05-96\\ RAL-TR-96-091\\ SHEP-96-29
\end{flushright}
\vspace{.2in}
\begin{center}
{\large{\bf \sheptitle}}
\bigskip \\ \shepauthor \\ \mbox{} \\ {\it \shepaddress} \\ \vspace{.5in}
{\bf Abstract} \bigskip 
\end{center} 
\setcounter{page}{0}
\shepabstract
\end{titlepage}

\section{Introduction \label{introduction}}

Over recent years there has been a good deal of activity
concerned with understanding the pattern of fermion masses and mixing angles
within the framework of supersymmetry and unification
(see next section for a review). The starting point of these analyses
is the idea that at high energies the Yukawa matrices exhibit a
degree of simplicity, typically involving texture zeroes,
which can be understood as resulting from some symmetry.
The types of symmetry which have been considered include
grand unified symmetry to account for the vertical mass splittings
within a family, and family symmetry to account for the
horizontal mass splittings between families. In order to restrict the rather
{\it ad hoc} \/nature of such models, one may appeal to a rigid theoretical
structure such as string theory in terms of which the high energy field
theory may be viewed as an effective low energy supergravity model
valid just below the string scale. Viewed from this perspective
certain classes of unified gauge group and family symmetry
appear to be more promising than others, and in addition one may
hope to begin to derive the entries of the Yukawa matrices
as low energy non-renormalisable operators which arise from the
string theory. 

In this paper, guided by the principles outlined in the previous paragraph,
we investigate the origin of Yukawa textures in
a class of  models based on the Pati-Salam $SU(4)\times SU(2)_L
\times SU(2)_R$ symmetry with gauged $U(1)$ family symmetries.

We shall follow both a bottom-up approach, in which the 
successful textures may be extracted from the known quark and lepton
masses and quark mixing angles, {\em and} \/a top-down approach in which
we shall begin to see how the desired operators may emerge from
a particular superstring construction. This  model
involves both quark-lepton unification, which leads to
Clebsch relations to describe the mass relations
within a particular family, and a
$U(1)_X$ gauged family symmetry which may account for family hierarchies.
Thus we are led to {\em vertical} \/and {\em horizontal} \/texture zeroes
which are a feature of this model. In the earlier parts of the paper
we shall focus on something we call the string-inspired $SU(4)\times O(4)$ 
 ($\sim SU(4)\otimes SU(2)_L \otimes SU(2)_R$) model which contains
 many of the features of a realistic string model
such as small group representations and a $U(1)_X$ family symmetry.
Within this simplified model we shall relate the high energy textures
to the low energy quark and lepton masses and quark mixing angles,
and so determine by a bottom-up procedure the operators which
are likely to be relevant at high energies. Later on we shall
focus on a particular string construction from which we learn how
non-renormalisable operators may be generated from first principles.

The detailed layout of the paper is as follows:
In section~\ref{textures} we review some ideas concerning
Yukawa textures, and summarise recent progress in this area.
In section~\ref{model} 
we briefly review the string-inspired $SU(4)\times O(4)$ model.
In section~\ref{sec:operators} we
discuss symmetric textures in the above model.
In section~\ref{asymmetric} we discuss the non-symmetric textures.
In section~\ref{numerical} we perform a full numerical analysis of the
non-symmetric models.
In section~\ref{familysymmetry}
we review the $U(1)_X$ family symmetry approach to the  model,
and perform an analysis relevant for the full
(symmetric and non-symmetric) model. In the subsequent two sections 
we present a viable string construction of the model  and
indicate how the non-renormalisable operators may arise in the
specific string construction.  
Finally section~\ref{conclusion} concludes the paper.

\section{Yukawa Textures \label{textures}}

The pattern of quark and lepton masses and quark mixing angles
has for a long time been a subject of fascination for particle physicists.
In terms of the standard model, this pattern arises from
three by three complex Yukawa matrices (54 real parameters)
which result in nine real eigenvalues plus four real mixing parameters
(13 real quantities) which can be measured experimentally. 
In recent years the quark and lepton
masses and mixing angles have been measured with increasing
precision, and this trend is likely to continue in the future as 
lattice QCD calculations provide increasingly accurate estimates
and B-factories come on-line.
Theoretical progress is less certain, although there has been a steady
input of theoretical ideas over the years and in recent times there is 
an explosion of activity in the area of supersymmetric unified models.
This approach presumes that at very high energies close to the unification
scale,
the Yukawa matrices exhibit a degree of simplicity, with simple relations 
at high energy corrected by the effects of renormalisation group (RG)
running down to low energy. For example the classic prediction that
the bottom and tau Yukawa couplings are equal at the unification scale
can give the correct low energy bottom and tau masses, providing that
one assumes the RG equations of the minimal 
supersymmetric standard model (MSSM)\cite{btau}\footnote{The
next-to-MSSM (NMSSM) with an additional low energy gauge singlet
works just as well \cite{btau2}.}.
In the context of the MSSM it is even possible that
the top, bottom and tau Yukawa couplings are all approximately equal near the
unification scale \cite{Yuk}, since although this results in the top and
bottom Yukawa couplings being roughly the same at low energy, one
can account for the large top to bottom mass ratio by invoking a large
value of $\tan \beta$ defined as the
ratio of vacuum expectation values (VEVs) of the two Higgs doublets
of the MSSM. 

These successes with the third family relations are not
immediately generalisable to the lighter families. For the remainder
of the Yukawa matrices, additional ideas are required in order
to understand the rest of the spectrum. One such idea is that of
texture zeroes: the idea that the Yukawa matrices at the unification 
scale are rather sparse; for example the Fritzsch ansatz \cite{Fritzsch}.
Although the Fritzsch texture does not work for           
supersymmetric unified models, there are other textures which do, for
example the Georgi-Jarlskog (GJ) texture \cite{GJ}
for the down-type quark and lepton matrices:
\begin{equation}
\lambda^E = \left(\begin{array}{ccc}
0 & \lambda_{12} & 0 \\
\lambda_{21} & - 3 \lambda_{22} & 0 \\
0 & 0 & \lambda_{33} \\ \end{array}\right) 
,\ \ 
\lambda^D = \left(\begin{array}{ccc}
0 & \lambda_{12} & 0 \\
\lambda_{21} & \lambda_{22} & 0 \\
0 & 0 & \lambda_{33} \\ \end{array}\right).
\label{GJ}
\end{equation}
After diagonalisation this leads to
$\lambda_{\tau} = \lambda_b$,
$\lambda_{\mu} = 3\lambda_s$, 
$\lambda_e = \lambda_d/3$
at the scale $M_{GUT}$ which result in (approximately) 
successful predictions at low energy. Actually the factor of 3 in the 22
element above arises from group theory: it is a Clebsch factor coming from
the choice of Higgs fields coupling to this element.

It is observed that if we choose the upper two by two block
of the GJ texture 
to be symmetric, $\lambda_{12}=\lambda_{21}$, and if we can disregard
contributions from the up-type quark matrix, then we also have the successful 
mixing angle prediction
\begin{equation}
V_{us}=\sqrt{\lambda_d/\lambda_s}.
\end{equation}
The data therefore supports the idea of symmetric matrices, and a texture
zero in the 11 position. Motivated by the desire for maximal predictivity,
Ramond, Roberts and Ross (RRR) \cite{RRR} have made a survey of possible
symmetric textures which are both consistent with data and involve
the maximum number of texture zeroes. Assuming GJ relations for the 
leptons, RRR tabulated five possible 
solutions for the up-type and down-type Yukawa matrices.
We list them below for completeness:

Solution 1:
\begin{equation}
\lambda^U = \left(\begin{array}{ccc}
0 & \sqrt{2}\lambda^6 & 0 \\
\sqrt{2}\lambda^6 & \lambda^4 & 0 \\
0 & 0 & 1 \\ \end{array}\right) 
,\ \ 
\lambda^D = \left(\begin{array}{ccc}
0 & 2\lambda^4 & 0 \\
2\lambda^4 & 2\lambda^3 & 4\lambda^3 \\
0 & 4\lambda^3 & 1 \\ \end{array}\right) 
\end{equation}

Solution 2:
\begin{equation}
\lambda^U = \left(\begin{array}{ccc}
0 & \lambda^6 & 0 \\
\lambda^6 & 0 & \lambda^2 \\
0 & \lambda^2 & 1 \\ \end{array}\right) 
,\ \ 
\lambda^D = \left(\begin{array}{ccc}
0 & 2\lambda^4 & 0 \\
2\lambda^4 & 2\lambda^3 & 2\lambda^3 \\
0 & 2\lambda^3 & 1 \\ \end{array}\right) 
\end{equation}

Solution 3:
\begin{equation}
\lambda^U = \left(\begin{array}{ccc}
0 & 0 & \sqrt{2}\lambda^4 \\
0 & \lambda^4 & 0 \\
\sqrt{2}\lambda^4 & 0 &  1 \\ \end{array}\right) 
,\ \ 
\lambda^D = \left(\begin{array}{ccc}
0 & 2\lambda^4 & 0 \\
2\lambda^4 & 2\lambda^3 & 4\lambda^3 \\
0 & 4\lambda^3 & 1 \\ \end{array}\right) 
\end{equation}

Solution 4:
\begin{equation}
\lambda^U = \left(\begin{array}{ccc}
0 & \sqrt{2}\lambda^6 & 0 \\
\sqrt{2}\lambda^6 & \sqrt{3}\lambda^4 & \lambda^2 \\
0 & \lambda^2 & 1 \\ \end{array}\right) 
,\ \ 
\lambda^D = \left(\begin{array}{ccc}
0 & 2\lambda^4 & 0 \\
2\lambda^4 & 2\lambda^3 & 0 \\
0 & 0 & 1 \\ \end{array}\right) 
\end{equation}

Solution 5:
\begin{equation}
\lambda^U = \left(\begin{array}{ccc}
0 & 0 & \lambda^4 \\
0 & \sqrt{2}\lambda^4 & \lambda^2/\sqrt{2} \\
\lambda^4 & \lambda^2/\sqrt{2} & 1 \\ \end{array}\right) 
,\ \ 
\lambda^D = \left(\begin{array}{ccc}
0 & 2\lambda^4 & 0 \\
2\lambda^4 & 2\lambda^3 & 0 \\
0 & 0 & 1 \\ \end{array}\right) 
\end{equation}
where $\lambda=0.22$, and the top and bottom Yukawa couplings
have been factored out for simplicity.
These textures are valid at the unification scale.
All of the solutions involve texture zeroes in the 11 entry.
Solutions 1,2, and 4 involve additional
texture zeroes in the 13=31 positions which are common to both up-type
and down-type matrices. Solutions 3 and 5 have no
texture zeroes which are common to both up-type and down-type matrices,
apart from the 11 entry. Thus solutions 1,2 and 4 involve
rather similar up-type and down-type matrices, while solutions
3 and 5 involve very different textures for the two matrices.

Having identified successful textures\footnote{Over the recent years,
there has been an extensive study of fermion mass 
matrices with zero textures \cite{oth1}.
}, the obvious questions are:
what is the origin of the texture zeroes? and: what is the origin
of the hierarchies (powers of the expansion parameter $\lambda$)?
A natural answer to both these questions was provided early on
by Froggatt and Nielsen (FN) \cite{FN}. The basic idea involves a
high energy scale $M$, a family symmetry group $G$,
and some new heavy matter of mass $M$ which transforms under $G$.
The new heavy matter
consists of some Higgs fields which are singlets under the vertical
gauge symmetry but non-singlets under $G$.
These break the symmetry $G$ by developing VEVs
$V$ smaller than the high energy scale, 
There are also some heavy fields which exist in vector-like
representations 
of the standard gauge group. The vector-like matter couples to ordinary
matter (quarks, leptons, Higgs) via the singlet Higgs, leading to
``spaghetti-like'' tree-level diagrams. Below the scale $V$
the spaghetti diagrams yield effective non-renormalisable 
operators which take the form of Yukawa couplings suppressed 
by powers of $\lambda =V/M$. In this way the hierarchies in the
Yukawa matrices may be explained, and the texture zeroes correspond to
high powers of $\lambda$. 

A specific realisation of the FN idea was 
provided by Ibanez and Ross (IR) \cite{IR}, based on the MSSM extended
by a gauged family $U(1)_X$ symmetry with $\theta$ and $\bar{\theta}$
singlet fields with opposite $X$ charges, plus new heavy Higgs fields
in vector representations\footnote{The generalisation to include
neutrino masses
is straightforward \cite{DLLRS}.
}.
Anomaly cancellation 
occurs via a Green-Schwarz-Witten
(GSW) mechanism, and the $U(1)_X$ symmetry is broken not far below the
string scale \cite{IR}. By making certain symmetric charge assignments,
IR showed that the RRR texture solution 2
could be approximately reproduced. To be specific, for a certain choice
of $U(1)_X$ charge assignments, IR generated Yukawa matrices of the
form:
\begin{equation}
\lambda^U = \left(\begin{array}{ccc}
\epsilon^8 & \epsilon^3 & \epsilon^4 \\
\epsilon^3 & \epsilon^2 & \epsilon \\
\epsilon^4 & \epsilon & 1 \\ \end{array}\right) 
,\ \ 
\lambda^D = \left(\begin{array}{ccc}
\bar{\epsilon}^8 & \bar{\epsilon}^3 & \bar{\epsilon}^4 \\
\bar{\epsilon}^3 & \bar{\epsilon}^2 & \bar{\epsilon} \\
\bar{\epsilon}^4 & \bar{\epsilon} & 1 \\ \end{array}\right) 
,\ \ 
\lambda^E = \left(\begin{array}{ccc}
\bar{\epsilon}^5 & \bar{\epsilon}^3 & 0 \\
\bar{\epsilon}^3 & \bar{\epsilon} & 0 \\
0  & 0  & 1 \\ \end{array}\right) 
\label{IR}
\end{equation}
These are symmetric in the expansion parameters $\epsilon$ and
$\bar{\epsilon}$, which are regarded as independent parameters.
This provides a neat and predictive framework, however there are
some open issues.
Although the order of the entries is fixed by the expansion parameters, 
there are additional
parameters of order unity multiplying each entry, making precise predictions
difficult. A way to address the problem of the unknown
coefficients has been proposed in
\cite{GRfp} where it has been shown that the various coefficients
may arise as a result of the infra-red fixed-point structure
of the theory beyond the Standard Model.

Note that the textures for up-type and down-type matrices
are of similar form, although the expansion parameters differ.
Also note that there are no true texture zeroes in the quark sector,
merely high powers
of the expansion parameter. Thus this example most closely resembles
RRR solution 2 with approximate texture zeroes in the 11 and 13=31 positions.
However, without the inclusion of 
coefficients, 
the identification is not exact.
The best fit to RRR solution 2 is obtained for
the identification 
$\epsilon \equiv  \lambda^2$, 
$\bar{\epsilon} \equiv \lambda$
(alternative identifications, like
$\epsilon \equiv  \lambda^2$, 
$\bar{\epsilon} \equiv 2\lambda^3$
lead to larger deviations).
However even this choice does not exactly correspond to RRR solution 2,
as can be shown by taking solution 2 and inserting the
numerical values of the entries:
\begin{equation}
\lambda^U = \left(\begin{array}{ccc}
0 & 1\times 10^{-4}& 0 \\
1\times 10^{-4} & 0 & 5\times 10^{-2} \\
0 & 5\times 10^{-2}  & 1 \\ \end{array}\right),\ 
\lambda^D = \left(\begin{array}{ccc}
0 & 5\times 10^{-3}& 0 \\
5\times 10^{-3} & 2\times 10^{-2} & 2\times 10^{-2} \\
0 & 2\times 10^{-2}& 1 \\ \end{array}\right) 
\label{RRRnumerical}
\end{equation}
We compare these numbers to the order of magnitudes
predicted by the symmetry argument,
making the identifications 
$\epsilon \equiv  \lambda^2$,
$\bar{\epsilon} \equiv \lambda$ 
\begin{equation}
\lambda^U = \left(\begin{array}{ccc}
3\times 10^{-11}& 1\times 10^{-4}& 5\times 10^{-6}\\
1\times 10^{-4} & 2\times 10^{-3}& 5\times 10^{-2} \\
5\times 10^{-6}& 5\times 10^{-2}  & 1 \\ \end{array}\right) 
,\  
\lambda^D = \left(\begin{array}{ccc}
5\times 10^{-6}& 1\times 10^{-2}& 2\times 10^{-3}\\
1\times 10^{-2} & 5\times 10^{-2}& 2\times 10^{-1} \\
2\times 10^{-3}& 2\times 10^{-1}  & 1 \\ \end{array}\right) 
\label{IRnumerical2}
\end{equation} 
Comparison of Eq.\ref{RRRnumerical} to Eq.\ref{IRnumerical2}
shows that while $\lambda^U$ is in good agreement, 
$\lambda^D$ differs. 
In Eq.\ref{IRnumerical2}, the $23=32$ element is an order of magnitude too
large. When the unknown couplings and phases are inserted
the scheme can be made 
to work. 
However, some tuning of the unknown parameters is implicit.
This can be avoided 
by introducing a small parameter $\delta$ into all the elements
apart from the 33 renormalisable element, so that Eq.\ref{IR} gets
replaced by\footnote{In our scheme we will have a Unified Yukawa matrix.
This, as we are going to see, will imply a common
expansion parameter for the up and down-type mass matrices
and the presence of a factor $\delta$ in the up-quark
mass matrix as well.}
\begin{equation}
\lambda^U = \left(\begin{array}{ccc}
\epsilon^8 & \epsilon^3 & \epsilon^4 \\
\epsilon^3 & \epsilon^2 & \epsilon \\
\epsilon^4 & \epsilon & 1 \\ \end{array}\right) 
,\ \ 
\lambda^D = \left(\begin{array}{ccc}
\delta\bar{\epsilon}^8 & \delta\bar{\epsilon}^3 & \delta\bar{\epsilon}^4 \\
\delta\bar{\epsilon}^3 & \delta\bar{\epsilon}^2 & \delta\bar{\epsilon} \\
\delta\bar{\epsilon}^4 & \delta\bar{\epsilon} & 1 \\ \end{array}\right) 
\label{deltaIR}
\end{equation}
The idea is that the suppression factor $\delta$ originates from some 
flavour independent physics, while the parameters
$\epsilon$ and 
$\bar{\epsilon}$ control the flavour structure of the
matrices. For example, suppose we take 
$\bar{\epsilon}\equiv \lambda$ as in the previous example
but scale down the entries by a factor of
$\delta = 0.2$. Then we would have,
\begin{equation}
\lambda^D = \left(\begin{array}{ccc}
1\times 10^{-6} & 2\times 10^{-3}& 4\times 10^{-4} \\
2\times 10^{-3} & 1\times 10^{-2} & 4\times 10^{-2} \\
4\times 10^{-4} & 4\times 10^{-2}& 1 \\ \end{array}\right) 
\label{deltaIRnumerical}
\end{equation} 
which provides a better description of the numerical values
required by the RRR analysis for solution 2 in Eq.\ref{RRRnumerical},
at the expense of introducing the parameter $\delta$.
This example indicates that if
family symmetries are to give the correct order of magnitude understanding
of Yukawa textures without any tuning of parameters,
then an extra parameter $\delta$ needs to be introduced as above.

Another aspect of the fermion mass spectrum that one would
like to understand, is that of the 
mass splitting within a particular family. For example the 
GJ texture in Eq.\ref{GJ} provides an understanding of the 
relationship between the charged lepton and down-type quark
Yukawa couplings within a given family, and in the 
simplest $U(1)_X$ scheme 
such relations are either absent or accidental, as seen 
in Eq.\ref{IR} where the form of 
$\lambda^E$ has been fixed by a parameter choice.
Unless such parameters are predicted by the theory, as in the
extension of the
initial IR scheme that is discussed in \cite{GRfp},
the only antidote is extra unification.
Then, the leptons share a representation with the quarks,
and the magic GJ factors of three originate from the fact that the
quarks have three colours. For example the $SO(10)$ model of 
Anderson et al \cite{Andersonetal} (with both low energy Higgs doublets
unified into a single {\underline{10}} representation)
predicts Yukawa unification
for the third family, GJ relations for the charged leptons and
down-type masses, and other Clebsch relations involving up-type
quarks. As in the IR approach, the approach followed by Anderson et al
is based on the FN ideas discussed above. Thus for example, only the third  
family is allowed to receive mass from the renormalisable operators in
the superpotential.  The remaining masses and mixings are generated 
from a minimal set of just three specially chosen non-renormalisable
operators whose coefficients are suppressed by a set of
large scales.
The 12=21 operator of Anderson et al is suppressed by
the ratio $(45_1/M)^6$, while the 23=32 and operators
are suppressed by $(45_{B-L}/45_1)^2$ and $(45_{B-L}S/45_1^2)$
where the 45's are heavy Higgs representations. In a complicated
multi-scale model such as this, the hierarchies between different
families are not understood in terms of a family symmetry 
such as the the $U(1)_X$ of IR\@. Indeed it is difficult to implement
a family symmetry in this particular scheme, as the latest attempts
based on global $U(2)$ \cite{BH} show. To be embedded into a string model,
GUTs such as $SO(10)$ require $k>1$ Kac-Moody levels.
With these higher Kac-Moody levels, simple orbifold
compactifications in which candidate gauge $U(1)_X$ family symmetries
are present do not easily emerge. Nevertheless there has been some
progress in this direction and three family $SO(10)$ and $E_6$ string-derived
models
have recently been classified \cite{SO10}.
Here we restrict our discussion to string constructions 
based on the simpler $k=1$ level of Kac-Moody algebras,
which are more ``string friendly''.

The $SU(4)\otimes O(4)$ string model can be viewed as the
simplest string-friendly unified extension of the standard model
which can lead to Clebsch relations of the kind we desire.
The Pati-Salam gauge group \cite{pati} 
may be broken without adjoint representations
and was considered as a unified string model
\cite{leo1},\cite{leo2} some time ago. This 
model has recently been the subject 
of renewed interest from the point of view of fermion masses \cite{422},
and an operator analysis has shown that it is possible to obtain
desirable features such as Yukawa unification for the third family,
and GJ type relations within this simpler model. A particular feature
of the published scheme which we would like to emphasise here is the idea of
{\em Clebsch texture zeroes} \/which arise from the group theory
of the Pati-Salam gauge group. These Clebsch zeroes were used to account for
the lightness of the up quark compared to the down quark, for example
\cite{422}. However the operator
analysis of \cite{422} did not address the question of the hierarchy
between families (no family symmetry was introduced for example),
nor the question of the origin of the
non-renormalisable operators. Here we shall introduce a $U(1)_X$ gauge
symmetry into the model
and combine it with the Clebsch relations previously used,
to provide a predictive scheme of fermion masses and mixing angles.
We shall also ensure that we obtain the correct order of magnitude
for all the entries of the Yukawa matrices from the symmetry breaking 
parameter, using structures like that of  Eq.\ref{deltaIR}. In our
case the quantity
$\delta$ will be identified with a bilinear of heavy Higgs fields
which are responsible for generating the Clebsch structures,
while the parameters such as $\epsilon$ will have trivial Clebsch structure
(singlets under the vertical gauge group) but will generate family hierarchies
from the flavour symmetry. This corresponds to there being two types
of heavy Higgs fields: Pati-Salam gauge singlets (corresponding to
IR $\theta$ and $\bar{\theta}$ fields)
which break the $U(1)_X$ family gauge group but leave the Pati-Salam
group unbroken, and $H,\bar{H}$ breaking fields whose bilinear forms are
$U(1)_X$ singlets but transform non-trivially under the Pati-Salam
gauge group, thereby giving interesting Clebsch structures.
The non-renormalisable operators of interest must therefore involve both
types of Higgs fields simultaneously. 
In view of the unusual nature of such operators, we shall provide a
string-based discussion of the origin of such operators.

It is worth emphasising that
the main features of the previous analysis, (like the assumption
of $U(1)$ symmetries, the introduction of singlet fields etc) appear
naturally in most of the recent string constructions. Therefore,
in the final sections of this paper we will try
to embed our analysis
in the context of realistic string models
which are constructed
within the free fermionic formulation \cite{fc}
of the heterotic string.
In doing so, we should keep in mind that,
in realistic  string constructions
\cite{fc,strmod,fermod}
there are usually
many constraints and in
general the resulting field theory is quite complicated.
Moreover, (in the language of the fermionic strings\cite{fermod}),
within the same choice of boundary conditions on the string  basis vectors 
of the world-sheet fermions, there
are numerous consistent  choices of the projection coefficients
which result in different Yukawa couplings 
multiplets and the large number of singlet fields which
are usually present. 
For this reason we shall try to develop a `string model' - independent
approach and  begin by considering a  field theory $SU(4)\times
 O(4)$  model, which possesses  the salient  features
of a realistic string model and at the same time is simpler to work with.

\section{The String Inspired $SU(4)\otimes O(4)$  Model \label{model}}
Here we briefly summarise the parts of the model which are relevant
for our analysis.  For a more complete discussion see \cite{leo1}. 
The gauge group is $SU(4)\otimes O(4)$, or equivalently 
\begin{equation}   
\mbox{SU(4)}\otimes \mbox{SU(2)}_L \otimes \mbox{SU(2)}_R. \label{422}
\end{equation}
The left-handed quarks and leptons are accommodated in the following
representations,
\begin{equation}
{F^i}^{\alpha a}=(4,2,1)=
\left(\begin{array}{cccc}
u^R & u^B & u^G & \nu \\ d^R & d^B & d^G & e^-
\end{array} \right)^i
\end{equation}
\begin{equation}
{\bar{F}}_{x \alpha}^i=(\bar{4},1,\bar{2})=
\left(\begin{array}{cccc}
\bar{d}^R & \bar{d}^B & \bar{d}^G & e^+  \\
\bar{u}^R & \bar{u}^B & \bar{u}^G & \bar{\nu}
\end{array} \right)^i
\end{equation}
where $\alpha=1,\ldots ,4$ is an SU(4) index, $a,x=1,2$ are
SU(2)$_{L,R}$ indices, and $i=1,2,3$ is a family index.  The Higgs
fields are contained in the following representations,
\begin{equation}
h_{a}^x=(1,\bar{2},2)=
\left(\begin{array}{cc}
  {h_2}^+ & {h_1}^0 \\ {h_2}^0 & {h_1}^- \\
\end{array} \right) \label{h}
\end{equation}
(where $h_1$ and $h_2$ are the low energy Higgs superfields associated
with the MSSM.) The two heavy Higgs representations are
\begin{equation}
{H}^{\alpha b}=(4,1,2)=
\left(\begin{array}{cccc}
u_H^R & u_H^B & u_H^G & \nu_H \\ d_H^R & d_H^B & d_H^G & e_H^-
\end{array} \right) \label{H}
\end{equation}
and
\begin{equation}
{\bar{H}}_{\alpha x}=(\bar{4},1,\bar{2})=
\left(\begin{array}{cccc}
\bar{d}_H^R & \bar{d}_H^B & \bar{d}_H^G & e_H^+ \\
\bar{u}_H^R & \bar{u}_H^B & \bar{u}_H^G & \bar{\nu}_H
\end{array} \right). \label{barH}
\end{equation}
The Higgs fields are assumed to develop VEVs,
\begin{equation}
<H>=<\tilde{\nu}_H>\sim M_{GUT}, \ \ 
<\bar{H}>=<\tilde{\bar{\nu}}_H>\sim M_{GUT}
\label{HVEV}  
\end{equation}
leading to the symmetry breaking at $M_{GUT}$
\begin{equation}
\mbox{SU(4)}\otimes \mbox{SU(2)}_L \otimes \mbox{SU(2)}_R
\longrightarrow
\mbox{SU(3)}_C \otimes \mbox{SU(2)}_L \otimes \mbox{U(1)}_Y
\label{422to321}
\end{equation}
in the usual notation.  
Under the symmetry breaking in Eq.\ref{422to321},
the bidoublet Higgs field $h$ in Eq.\ref{h} splits into two
Higgs doublets $h_1$, $h_2$ whose neutral components subsequently
develop weak scale VEVs,
\begin{equation}
<h_1^0>=v_1, \ \ <h_2^0>=v_2 \label{vevs1}
\end{equation}
with $\tan \beta \equiv v_2/v_1$.

In addition to the Higgs fields in Eqs.~\ref{H},\ref{barH} the model
also involves an SU(4) sextet field $D=(6,1,1)$ and four 
singlets $\phi_0$ and $\varphi_i$, $i= 1,2,3$. $\phi_0$ is going to
acquire an electroweak VEV in order to
realise the electroweak higgs mixing, while $\varphi_i$
will participate in an extended `see-saw'
mechanism to obtain light majorana masses for the left - handed 
neutrinos. Under the symmetry property
$\varphi_{1,2,3}\ra (-1)\times \varphi_{1,2,3}$ 
and $H (\bar{H})\ra (-1)\times H (\bar{H})$ the tree level mass terms of
the superpotential of the model read \cite{leo1}:
\begin{equation}
W =\lambda^{ij}_1F_i\bar{F}_jh
+\lambda_2HHD+\lambda_3\bar{H}\bar{H}D+\lambda^{ij}_4H\bar{F_j}
\varphi_i +\mu\varphi_i\varphi_j+\mu hh \label{W}
\end{equation}
where $\mu = <\phi_0>\sim {\cal O}(m_W)$. 
The last term generates the higgs mixing between the two 
SM higgs doublets in order to prevent the appearance of a massless
electroweak axion. Note that we have banned terms which might lead to
unacceptably large neutrino-higgsino mixing~\cite{fp}.
The superpotential Eq.\ref{W} leads to the following neutrino mass matrix
\cite{leo1}
\begin{eqnarray}
{\cal M}_{\nu,N^c,\varphi} = \left(\begin{array}{ccc}
0 & m^{ij}_u & 0 \\
m^{ji}_u& 0 & M_{GUT} \\
0  & M_{GUT}  & \mu \\ \end{array}\right) 
\label{RRRioperators}
\end{eqnarray}
in the basis $(\nu_i , \bar{\nu}_j ,\varphi_k)$. Diagonalisation of the above
gives three light neutrinos with masses of the order $(m_u^{ij})^2/M_{GUT}$
as required, and leaves right handed majorana masses of the order $M_{GUT}$.
Additional terms not included in Eq.\ref{W} may be forbidden by imposing 
suitable discrete or continuous symmetries, the details of which
need not
concern us here.  The $D$ field carries colour and therefore 
 not develop  a VEV but the terms in Eq.\ref{W}
$HHD$ and $\bar{H} \bar{H}D$ combine the  colour triplet parts of $H$, $\bar{H}$
and $D$ into acceptable GUT-scale mass terms \cite{leo1}. When the $H$ fields
attain their VEVs at $M_{GUT}\sim10^{16}$ GeV, the superpotential of Eq.\ref{W}
reduces to that of the MSSM augmented by neutrino masses. Below $M_{GUT}$ the 
part of the superpotential involving matter superfields is just
\begin{equation}
W =\lambda^{ij}_UQ_i\bar{U}_jh_2+\lambda^{ij}_DQ_i\bar{D}_jh_1
+\lambda^{ij}_EL_i\bar{E}_jh_1+ \lambda^{ij}_NL_i\bar{\nu}_jh_2 + \ldots
\label{MSSMmatter}
\end{equation}
The Yukawa couplings in Eq.\ref{MSSMmatter} satisfy the
boundary conditions
\begin{equation}   
\lambda^{ij}_1 (M_{GUT}) \equiv \lambda^{ij}_U(M_{GUT}) 
= \lambda^{ij}_D (M_{GUT})=
\lambda^{ij}_E(M_{GUT}) = \lambda^{ij}_N(M_{GUT}).
 \label{boundary}
\end{equation}
Thus, Eq.(\ref{boundary}) retains the successful relation
$m_{\tau}=m_b$ at $M_{GUT}$. Moreover from the
relation $\lambda^{ij}_U(M_{GUT}) = \lambda^{ij}_N(M_{GUT})$,
and the fourth term in Eq.\ref{W}, we obtain through the see-saw 
mechanism light neutrino masses $\sim {\cal O}(m_u^2/M_{GUT})$
which satisfy the experimental limits.

\section{Symmetric Textures \label{sec:operators}}

In this section we briefly review the results of the operator analysis of
ref.\cite{422}, then introduce our new approach based on new operators. 
We discuss the RRR textures as a simple example of the new method.

The boundary conditions listed in Eq.\ref{boundary} lead to
unacceptable mass relations for the light two families. Also, the
large family hierarchy in the Yukawa couplings appears to be unnatural
since one would naively expect the dimensionless couplings all to be  
of the same order. This leads us to the conclusion that the
$\lambda^{ij}_1$ in Eq.\ref{W} may not originate from the usual
renormalisable tree level dimensionless coupling.  We allow a  
renormalisable Yukawa coupling in the 33 term only and generate the
rest of the effective Yukawa couplings by non-renormalisable operators
that are suppressed by some higher mass scale. This suppression
provides an explanation for the observed fermion mass hierarchy.

In ref.\cite{422} we restricted ourselves to all possible non-renormalisable
operators which can be constructed from different group theoretical
contractions of the fields:
\begin{equation}
O_{ij}\sim (F_i\bar{F}_j
)h\left(\frac{H\bar{H}}{{M}^2}\right)^n+{\mbox H.c.} \label{op}
\end{equation}
where we have used the fields $H,\bar{H}$ in Eqs.\ref{H},\ref{barH}
and $M$ is the large scale $M>M_{X}$.  The idea is that when $H,
\bar{H}$ develop their VEVs, such operators will become effective
Yukawa couplings of the form $h F \bar{F}$ with a small coefficient of
order $M_{GUT}^2/M^2$.  We considered up to $n=2$ operators.
The motivation for using $n=2$ 
operators is simply that such higher dimension operators
are generally expected to lead to smaller effective couplings more
suited to the 12 and 21 Yukawa entries. However, in our field theory 
approach we shall restrict ourselves to the simple case considering 
only $n=1$ operators with the required suppression factors originating
from a separate flavour sector. 
We will leave the question of the definite origin of the operators 
for now. Instead we merely note that one could
construct a FN sector to motivate the operators, or that one might expect
such operators to come directly out of a string theory.
In section~\ref{familysymmetry}
we shall introduce a $U(1)_X$ family symmetry into the model,
which is broken at a scale $M_X>M_{GUT}$ by the VEVs of the Pati-Salam
singlet fields $\theta$ and $\bar{\theta}$. According to the ideas discussed
in section~\ref{textures} we shall henceforth consider operators of the
form
\begin{equation} \label{newop}
O_{ij}\sim (F_i\bar{F}_j)h\left(\frac{H\bar{H}}{M^2}\right)
\left(\frac{\theta^n \bar{\theta}^m}{{M'}^{n+m}}\right)+
{\mbox h.c.}
\label{newoperators} 
\end{equation}
where $M'$ represents a high scale $M' > M_{GUT}$ which may be identified 
either with the $U(1)_X$ breaking scale $M_X$ or with the string scale. We have
further assumed the form of the operators in Eq.\ref{op}  corresponding to $n=1$
and  glued onto these operators  arbitrary powers of the singlet fields $\theta ,
\bar{\theta}$.  Note that the single power of $(H\bar{H})$ is present in 
every entry of the matrix and plays the role 
of the factor of $\delta$ in Eq.\ref{deltaIR}.
However, unlike the previous factor of $\delta$, the factor of
$(H\bar{H})$ here carries important group theoretical Clebsch information.
In fact Eq.\ref{newop} amounts to assuming a sort of {\em
factorisation} \/of 
the operators with the family hierarchies being completely 
controlled by the $\theta , \bar{\theta}$ fields as in IR\@,
with $m,n$ being dependent on $i,j$, and the horizontal 
splittings being controlled by the Clebsch factors in $(H\bar{H})$.
However this factorisation is not complete since
we shall assume that the Clebsch factors 
have a family dependence, i.e.\ they depend on $i,j$. 
We select the Clebsch factor in each entry  from phenomenological
arguments.

\begin{table} 
\begin{center}
\begin{tabular}{|c|c|c|c|c|} \hline
 & $Q \bar{U} h_2$ & $Q \bar{D} h_1$ & $L \bar{E} h_1$ & $L \bar{N}
h_2$
\\ \hline
$O^A$ &1 & 1 & 1 &1 \\ 
$O^B$ &1 & -1& -1 &1 \\ 
$O^C$ &$\frac{1}{\sqrt{5}}$ & $\frac{1}{\sqrt{5}}$ 
& $\frac{-3}{\sqrt{5}}$ &$\frac{-3}{\sqrt{5}}$  \\ 
$O^D$ &$\frac{1}{\sqrt{5}}$ & $\frac{-1}{\sqrt{5}}$
& $\frac{3}{\sqrt{5}}$ & $\frac{-3}{\sqrt{5}}$\\ 
$O^G$ & 0 & $\frac{2}{\sqrt{5}}$ & $\frac{4}{\sqrt{5}}$ & 0 \\
 $O^H$ & 4/5 & 2/5 & 4/5 & 8/5 \\
$O^K$ & 8/5 & 0 & 0 & 6/5 \\
$O^M$ &0 & $\sqrt{2}$ & $\sqrt{2}$ &0 \\ 
$O^N$ &2&0&0&0\\ 
$O^R$ &0&$\frac{8}{5}$& $\frac{6}{5}$& 0 \\
$O^W$ &0 & $\sqrt{\frac{2}{5}}$& -3$\sqrt{\frac{2}{5}}$&0\\
$O^S$ & $\frac{8}{5 \sqrt{5}}$ & $\frac{16}{5 \sqrt{5}}$ &
$\frac{12}{5 \sqrt{5}}$ & $\frac{6}{5 \sqrt{5}}$ \\
\hline
\end{tabular}
\end{center}
\label{tab:subset}
\caption{{\small When the Higgs fields develop their VEVs at $M_{GUT}$, the
$n=1$ operators utilised lead to the effective Yukawa couplings with
Clebsch coefficients as shown. We have included the relative
normalisation for each of the operators.
The full set of $n=1$ operators and Clebsch coefficients is given in Appendix 1.
These $n=1$ operators were used in the lower right hand block of the
Yukawa matrices in the analysis of 
ref.~\protect\cite{422}.}}
\end{table}

As a first example of our new approach we shall consider the RRR textures
discussed in section~\ref{textures}. Our first observation is that,
restricting ourselves to $n=1$ operators, there are no large Clebsch
ratios between the up-type and down-type quarks for any of the operators.
This means that it is very difficult to reproduce RRR solutions such as
solution 2 where the 12 element of the down-type matrix in 
Eq.\ref{RRRnumerical}, for example, is 50 times larger than its
up-type counterpart. Of course this can be achieved by requiring an
accurate cancellation between two operators, but such a tuning of coefficients
looks ugly and unnatural, and we reject it. On the other hand the
$n=1$ Clebsch coefficients in Table 1 include examples of {\em zero Clebsch coefficients},
where the  contribution to the up-type matrix, for example, is
precisely zero. Similarly there are {\em zero Clebsch coefficients} 
\/for the down-type quarks (and charged leptons). 
The existence of such {\em zero Clebsch coefficients}
\/enables us to reproduce the RRR texture solutions 3 and 5 without
fine-tuning.
Interestingly they are precisely the solutions which are not possible
to obtain by the standard IR symmetry approach, which favours
solutions 1,2 and 4 and for which the up-type and down-type structures
are similar.
Thus our approach is capable of describing the RRR solutions which are
complementary to those described by the IR symmetry 
approach\footnote{In \cite{LLSV}, two of us used an alternative
approach in
order to reproduce the structure of
solutions 1 and 3 of RRR by the implementation of a symmetry. These
solutions were
found to lead to the optimal predictions for neutrino masses
and mixings. This has been achieved by a proper choice of charges
(integer/half-integer) and by imposing
residual $Z_2$ symmetries which forbid different entries
in the up and down-quark mass matrices.}.
To take a specific example let us begin by ignoring the flavour
dependent singlet fields, and 
consider the symmetric $n=1$ operator texture,
\begin{equation}
\lambda = \left(\begin{array}{ccc}
0 & O^M & O^{N} \\
O^M & O^W + s.d. & O^{N} \\
O^{N}  & O^{N}  & O_{33} \\ \end{array}\right) 
\label{RRR5operators}
\end{equation}
where $O_{33}$ is the renormalisable operator, ${s.d.}$ stands for a
sub-dominant operator with a suppression factor compared to the other
dominant operator in the same entry.
Putting in the Clebsch coefficients from Table 1 we arrive at the
component Yukawa matrices, at the GUT scale, of
\begin{equation}
\lambda^U = \left(\begin{array}{ccc}
0 & 0  & 2\lambda_{13}^U \\
0 & \lambda_{22}^U & 2\lambda_{23}^U \\
2\lambda_{13}^U  & 2\lambda_{23}^U   & 1 \\ \end{array}\right) 
\label{RRR5upcomponents}
\end{equation}
\begin{equation}
\lambda^D = \left(\begin{array}{ccc}
0 & \sqrt{2}\lambda_{12}^D & 0\\
\sqrt{2}\lambda_{12}^D & \lambda_{22}^D\sqrt{2}/\sqrt{5} & 0 \\
0  & 0    & 1 \\ \end{array}\right) 
\label{RRR5downcomponents}
\end{equation}
\begin{equation}
\lambda^E = \left(\begin{array}{ccc}
0 & \sqrt{2}\lambda_{12}^D & 0\\
\sqrt{2}\lambda_{12}^D & 3\lambda_{22}^D\sqrt{2}/\sqrt{5} & 0 \\
0  & 0    & 1 \\ \end{array}\right) 
\label{RRR5leptoncomponents}
\end{equation}
where $\lambda_{22}^D$ and $\lambda_{22}^E$ arise from the
dominant $O^W_{22}$ operator and $\lambda_{22}^U$ comes from a sub-dominant
operator that is relevant because of the texture zero Clebsch in the up
sector of $O^W_{22}$.
The zeroes in the matrices correspond to those of the RRR solution 5,
but of course in our case they arise from the Clebsch zeroes rather than
from a family symmetry reason. The numerical values corresponding to RRR
solution 5 with the correct phenomenology are,
\begin{equation}
\lambda^U = \left(\begin{array}{ccc}
0 & 0 & 2\times 10^{-3} \\
0 & 3\times 10^{-3} & 3\times 10^{-2} \\
2\times 10^{-3} & 3\times 10^{-2}  & 1 \\ \end{array}\right) 
,\ 
\lambda^D = \left(\begin{array}{ccc}
0 & 5\times 10^{-3}& 0 \\
5\times 10^{-3} & 2\times 10^{-2} & 0 \\
0 & 0 & 1 \\ \end{array}\right) 
\label{RRR5numerical}
\end{equation}
Thus, the hierarchy $\lambda_{22}^U << \lambda_{22}^D$ is explained by
a Clebsch zero and a suppression factor of the sub-dominant operator.
Using Eq.\ref{RRR5numerical}
we can read off the values of the couplings which 
roughly correspond to a unified matrix of dominant couplings
\begin{equation}
\lambda = \left(\begin{array}{ccc}
0 & 3\times 10^{-3}& 1\times 10^{-3} \\
3\times 10^{-3} & 2\times 10^{-2} & 2\times 10^{-2} \\
1\times 10^{-3} & 2\times 10^{-2} & 1 \\ \end{array}\right) 
\label{unifiednumerical}
\end{equation}
where we have extracted the Clebsch factors. We find it particularly
elegant that the whole quark and lepton spectrum is controlled
by a unified Yukawa matrix such as in Eq.\ref{unifiednumerical}
with all the vertical splittings controlled by Clebsch factors.

At this stage we could introduce a $U(1)_X$ symmetry of the IR kind,
and the flavour dependent singlet fields in order to account for the
horizontal family hierarchy of couplings in Eq.\ref{unifiednumerical}.
In the present case we must remember that there is
a small quantity $\delta$ multiplying every
non-renormalisable entry as in  Eq.\ref{deltaIR},
corresponding to the $n=1$
bilinear $\delta \equiv \frac{v\bar{v}}{M^2}$ which we have
required to be present in every non-renormalisable entry.
Thus we can understand Eq.\ref{unifiednumerical}
as resulting from a structure like,
\begin{equation}
\lambda = \left(\begin{array}{ccc}
\delta {\epsilon}^8 & \delta {\epsilon}^3 & \delta {\epsilon}^4 \\
\delta {\epsilon}^3 & \delta {\epsilon}^2 & \delta {\epsilon} \\
\delta {\epsilon}^4 & \delta {\epsilon} & 1 \\ \end{array}\right) 
\label{deltaIR2}
\end{equation}
where we identify $\epsilon \equiv \lambda = 0.22$
and set $\delta \approx 0.2$ 
which gives the correct orders of magnitude for the entries,
rather similar to the case we discussed in Eq.\ref{deltaIRnumerical}.
Here of course the considerations apply to the unified Yukawa matrix,
however, not just the down-type quark matrix. The details of the 
$U(1)_X$ family symmetry analysis are discussed in
section~\ref{familysymmetry}.
Here we simply note that such an analysis can lead to a structure
such as the one assumed in Eq.\ref{deltaIR2}.

A similar analysis could equally well be applied to RRR solution 3.
In both cases we are led to a pleasing scheme which involves no unnatural
tuning of elements, and naturally combines the effect of Clebsch coefficients
with that of family symmetry suppression, in a simple way.
The existence of the Clebsch texture zeroes thus permits
RRR solutions 3 and 5 which are impossible to obtain 
otherwise within the general framework presented here.

\section{Non-Symmetric  Textures \label{asymmetric}}

In this section we up-date the non-symmetric textures
based on both $n=1$ and $n=2$ operators introduced in ref.\cite{422},
then extend the new approach introduced in the previous section to the
non-symmetric domain. As in the previous section, 
we shall begin by ignoring the effect of the singlet fields,
which will be discussed in section~\ref{familysymmetry}.

As discussed in Appendix 2 we shall modify the analysis of 
Ref.\cite{422} to only include
the lower 2 by 2 block Ansatz:
\begin{equation}
A_1= \left[\begin{array}{cc}
O_{22}^W + s.d. & 0 \\
O_{32}^C & O_{33} \\ \end{array}\right].
\label{A_1}
\end{equation}
This is then combined with 
the upper 2 by 2 blocks considered in
ref.\cite{422}:
\begin{eqnarray} B_1 &=& \left[\begin{array}{cc} 0 &  O^1\\
O^{Ad} & X \end{array}\right]\label{suclighti}\\ B_2 &=&
\left[\begin{array}{cc} 0 &  O^2\\ O^{Ad} & X
\end{array}\right]\\ B_3 &=& \left[\begin{array}{cc} 0 & 
O^3\\ O^{Ad} & X \end{array}\right]\\ B_4 &=& \left[\begin{array}{cc}
0 &  O^1\\ O^{Dd} & X \end{array}\right]\\ B_5 &=&
\left[\begin{array}{cc} 0 &  O^2\\ O^{Dd} & X
\end{array}\right]\\ B_6 &=& \left[\begin{array}{cc} 0 & 
O^3\\ O^{Dd} & X \end{array}\right]\\ B_7 &=& \left[\begin{array}{cc}
0 &  O^1\\ O^{Md} & X \end{array}\right]\\ B_8 &=&
\left[\begin{array}{cc} 0 &  O^2\\ O^{Md} & X
\end{array}\right], \label{suclightii}
\end{eqnarray} where $X$ stands for whatever is left in the 22 position,
after the lower 2 by 2 submatrix has been diagonalised. 
The Clebsch coefficients of the $n=2$ operators used in
Eqs.\ref{suclighti}-\ref{suclightii} are displayed in
Table~\ref{tab:n2ops} but we refer the reader to ref.\cite{422} for
the explicit realisation of these operators in terms of the component
fields for reasons of brevity.
\begin{table}
\begin{center}
\begin{tabular}{|c|c|c|c|c|} 
\hline & $Q \bar{U} h_2$ & $Q \bar{D} h_1$
 & $L \bar{E} h_1$ & $L \bar{N} h_2$ \\
 \hline $O^{Ad}$ &$\frac{4 \sqrt{2}}{25}$ &$ \frac{12 \sqrt{2}}{25}$&
$\frac{9 \sqrt{2}}{25}$&$\frac{3 \sqrt{2}}{25}$\\ 
$O^{Dd}$ &$\frac{1}{\sqrt{5}}$ &$ \frac{3}{\sqrt{5}}$
&$\frac{3}{\sqrt{5}}$&$\frac{1}{\sqrt{5}}$ \\
 $O^{Md}$ &$\frac{\sqrt{2}}{5}$ & $\frac{3\sqrt{2}}{5}$&
$\frac{6 \sqrt{2}}{5}$&$\frac{2 \sqrt{2}}{5}$\\ 
$O^1$ &0 & $\sqrt{2}$ & $\sqrt{2}$&0 \\
 $O^2$ &0 & $\frac{8}{5}$ & $\frac{6}{5}$&0 \\
 $O^3$ &0 & $\frac{2}{\sqrt{5}}$ & $\frac{4}{\sqrt{5}}$&0 \\ \hline
\end{tabular}
\end{center}
\caption{{\small Clebsch coefficients of $n=2$ operators previously utilised.}
\label{tab:n2ops}}
\end{table}
The Ans\"atze listed above present problems because of the breakdown of 
matrix perturbation theory\footnote{When the magnitudes of
$H_{21},H_{12}$ and $H_{22}$ are calculated they are each of the same order
in the down Yukawa matrix, thus violating the hierarchy in
Eq.\ref{matrixform} that was assumed in the calculation of the
predictions.}. For purposes of comparison with the new scheme
involving only $n=1$ operators, we will recalculate the predictions
for each of
the models from ref.\cite{422} numerically in the next section.

We now turn our attention to the new approach introduced
in the previous section, based on $n=1$ operators together with
singlet fields which for the moment we shall ignore. In this case
the 21 operator used in ref.\cite{422} which gave an up Clebsch coefficient 
1/3 times smaller than the down Clebsch is not available if we only
use $n=1$ operators.
We must therefore use a combination of two operators in the 21
position that allow the up entry to be a bit smaller than the down
entry. We require that the combination provide a Clebsch relation
between $\lambda_{21}^D$ and $\lambda_{21}^E$ for predictivity. The
two operators cancel slightly in the up sector, but as shown later
this cancellation is $\sim O(1)$ and therefore acceptable.
The result of this is that the
prediction of $V_{ub}$ is lost; however this prediction was almost
excluded by experiment anyway, and a more accurate numerical estimate
which does not rely on matrix perturbation theory confirms that 
$V_{ub}$ in ref.\cite{422} is too large. So the loss of the
$V_{ub}$ prediction is to be welcomed! The Clebsch effect of the 12
operator (with a zero Clebsch for the up-type quarks) can easily
be reproduced at the $n=1$ level by the operator $O^M$ for example.

To get some feel for the procedure we will follow, we first discuss
a simple example of a non-symmetric texture, ignoring complex phases
for illustrative purposes. Restricting ourselves to $n=1$ operators,
we consider the lower block to be $A_1$ and the upper block to be
the modified texture as discussed in the previous paragraph.
Thus we have,
\begin{equation}
\lambda = \left(\begin{array}{ccc}
0 & O^M & 0 \\
O^M+O^A & O^W+s.d. & 0 \\
0  & O^C  & O_{33} \\ \end{array}\right) 
\label{nonsym}
\end{equation}
where $O_{33}$ is the renormalisable operator.
Putting in the Clebsch coefficients from Table~\ref{tab:subset}  we arrive at the
component Yukawa matrices, at the GUT scale, of
\begin{equation}
\lambda^U = \left(\begin{array}{ccc}
0 & 0  & 0 \\
\lambda_{21}^U & \lambda_{22}^U & 0 \\
0  & \sqrt{2}\lambda_{32}^U/\sqrt{5}   & 1 \\ \end{array}\right) 
\label{nonsymupcomponents}
\end{equation}
\begin{equation}
\lambda^D = \left(\begin{array}{ccc}
0 & \sqrt{2}\lambda_{12}^D & 0\\
\sqrt{2}\lambda_{21}^D & \lambda_{22}^D/\sqrt{5} & 0 \\
0  & -\sqrt{2}\lambda_{32}^U/\sqrt{5}    & 1 \\ \end{array}\right) 
\label{nonsymdowncomponents}
\end{equation}
\begin{equation}
\lambda^E = \left(\begin{array}{ccc}
0 & \sqrt{2}\lambda_{12}^D & 0\\
\sqrt{2}\lambda_{21}^D & 3\lambda_{22}^D/\sqrt{5} & 0 \\
0  & -3 \sqrt{2}\lambda_{32}^U/\sqrt{5}    & 1 \\ \end{array}\right) 
\label{nonsymleptoncomponents}
\end{equation}
where $\lambda_{22}^U$ and $\lambda_{22}^D$ arise from the
difference and sum of two operators whose normalisation factor
of $\sqrt{5}$ has been explicitly inserted, and similarly for
$\lambda_{21}^U$ and $\lambda_{21}^D$.
To obtain the numerical values of the entries we use some 
typical GUT-scale values of Yukawa couplings and 
CKM elements (see ref.\cite{422}) as follows:
\begin{equation}
\lambda_{33}=1, \lambda_c=0.002, \lambda_s=0.013, \lambda_{\mu}=0.04,
\lambda_u=10^{-6}, \lambda_d=0.0006, \lambda_e=0.0002,
\label{Yukvalues}
\end{equation}
\begin{equation}
V_{cb}=0.05,\, V_{us}=0.22,\, V_{ub}=0.004
\label{CKMvalues}
\end{equation}
where we have assumed,
\begin{equation}
\alpha_s=0.115,\, m_b=4.25,\, \tan\beta = 55, \, m_t=180 GeV
\label{lowparams}
\end{equation}
The textures in Eqs.\ref{nonsymupcomponents},~\ref{nonsymdowncomponents}
and~\ref{nonsymleptoncomponents} imply that the 22 eigenvalues are
just equal to the 22 elements (assuming matrix perturbation theory
is valid -- see later), and $\lambda_{32}^U=V_{cb}/2=0.025$. 
Thus we have $\lambda_{22}^U=0.004$,$\lambda_{22}^D=0.03$.
The remaining
parameters are determined from the relations,
\begin{equation}
\lambda_u=0,\, \lambda_d=3\lambda_e=\lambda_{21}^D\sqrt{2}\lambda_{12}^D 
/\lambda_s,\, V_{ub}=\lambda_{21}^UV_{cb}/\lambda_c
\label{nonsymprediction}
\end{equation}
Note that the up quark mass looks like it is zero, but
in practice we would expect some higher dimension operator to be
present which will give it a small non-zero value. 
We thus have three equations and three unknowns, and solving we find
$\lambda_{21}^U=2\times 10^{-4}$, 
$\lambda_{21}^D=2\times 10^{-3}$, 
$\lambda_{12}^D=3\times 10^{-3}$.
The difference between $\lambda_{21}^U$ and $\lambda_{21}^D$
requires suppression of $O^A$ caused by the Clebsch zero in the
dominant operator $O^M$.
Thus the unified Yukawa matrix involves operators with the following
approximate numerical coefficients,
\begin{equation}
\lambda = \left(\begin{array}{ccc}
0 & 3\times 10^{-3}& 0 \\
3\times 10^{-3} & 1.5\times 10^{-2} & 0 \\
0 & 2.5\times 10^{-2} & 1 \\ \end{array}\right) 
\label{nonsymunifiednumerical}
\end{equation}
where we have extracted the Clebsch factors, and the 22 and 21 values 
in Eq.\ref{nonsymunifiednumerical} refer
to each of the two operators in this position separately.
The numerical values in Eq.\ref{nonsymunifiednumerical}
are not dissimilar from those in Eq.\ref{unifiednumerical},
in particular the upper 2 by 2 block is
symmetrical with the same values as before.
In this case the lower 2 by 2 block has a texture zero in the 23 position,
as well as the 31 and 13 positions, but otherwise the numerical values
are very similar to those previously obtained in Eq.\ref{unifiednumerical}.
Thus this particular non-symmetric texture can be described
by a structure of the kind,
\begin{equation}
\lambda = \left(\begin{array}{ccc}
\delta {\epsilon}^{big}& \delta {\epsilon}^3 & \delta {\epsilon}^{big} \\
\delta {\epsilon}^3 & \delta {\epsilon}^{1or2} & \delta {\epsilon}^{big} \\
\delta {\epsilon}^{big} & \delta {\epsilon} & 1 \\ \end{array}\right) 
\label{deltanonsym}
\end{equation}
where we identify $\epsilon \equiv \lambda = 0.22$
and set $\delta \approx 0.1$ as before. Can such a structure for the
$\epsilon$'s be obtained from the $U(1)_X$ symmetry? This will be discussed
in section~\ref{familysymmetry}.

There is no reason to restrict ourselves to non-symmetric textures
with a zero in the 13 and 31 position, as assumed in ref.\cite{422}.
For example the following texture is also viable,
amounting to a hybrid of the symmetric case
considered in Eq.\ref{RRR5operators} and the non-symmetric lower block
 just considered.
\begin{equation}
\lambda = \left(\begin{array}{ccc}
0 & O^M & O^N \\
O^M & O^W & 0 \\
O^N  & O^C  & O_{33} \\ \end{array}\right) 
\label{hybridsym}
\end{equation}
Here, $O_{33}$ is the renormalisable operator.
We now perform a general operator analysis of the non-symmetric 
case, assuming $n=1$ operators for all non-zero entries (apart
from the 33 renormalisable entry). In this general analysis there are
two classes of texture: those with universal texture zeroes in the 13
and 31 position (essentially $n=1$ versions of the textures considered
in ref.\cite{422}) and new textures with  non-zero entries in the
13 and/or 31 position. For now we will not consider the cases with
operators in the 13 or 31 positions for reasons of brevity.
In the general analysis we repeat the above 
procedure, being careful about phases, and obtain some numerical
estimates of the magnitude of each entry which will be explained
in terms of the $U(1)_X$ family symmetry as discussed in the next section.

With the above discussion in mind, we consider the new scheme in which
the dominant operators in the Yukawa matrix are $O_{33}$, $O_{32}^C$,
$O_{22}^W$, $O_{21}, \tilde{O}_{21}$ and $O_{12}$, where the last 
three operators are left general and
will be specified later. We are aware from
the analysis in ref.\cite{422} that $O_{12}$ must have a zero Clebsch
coefficient in the up sector. A combination of two operators must
then provide a non-zero $O_{21}$ entry to provide a big enough
$V_{ub}$, an additional much more suppressed operator elsewhere in
the Yukawa matrix gives the up quark a small mass.
At $M_{GUT}$ therefore, the Yukawa matrices are of the form
\begin{equation}
\lambda^I = \left[ \begin{array}{ccc} 0 & H_{12} e^{i \phi_{12}}
x_{12}^I 
& 0
\\ H_{21} x_{21}^I e^{i \phi_{21}}+ \tilde{H}_{21} \tilde{x}_{21}^I
e^{i\tilde{\phi}_{21}} & H_{22} x_{22}^I e^{i \phi_{22}} &
0 \\
0 & H_{32} x_{32}^I e^{i \phi_{32}} & H_{33} e^{i \phi_{33}} \\
\end{array}\right], \label{dom}
\end{equation}
where only the dominant operators are listed. The $I$ superscript
labels the charge sector and $x_{ij}^I$ refers to the Clebsch
coefficient relevant to the charge sector $I$ in the $ij^{th}$
position. $\phi_{ij}$ are unknown phases and $H_{ij}$ is the magnitude
of the effective dimensionless Yukawa coupling in the $ij^{th}$
position. Any subdominant operators that we introduce will be denoted
below by a prime and it should be borne in mind that these will only
affect the up matrix. So far, the known Clebsch coefficients are
\begin{eqnarray}
x_{12}^U = 0 & & \nonumber \\
x_{22}^U = 0 & x_{22}^D = 1 & x_{22}^E = -3 \nonumber \\
x_{32}^U = 1 & x_{32}^D = -1 & x_{32}^E = -3.
\label{cleb}
\end{eqnarray}
We have just enough freedom in rotating the phases of $F_{1,2,3}$ and
$\bar{F}_{1,2,3}$ to get rid of all but one of the phases in
Eq.\ref{dom}.
When the subdominant operator is added, the Yukawa matrices are
\begin{eqnarray}
\lambda^U &=& \left[\begin{array}{ccc} 0 & 0 & 0\\
H_{21}^U e^{i \phi_{21}^U} & {H_{22}}' e^{i {\phi_{22}}'} & 0 \\
0 & H_{32} x_{32}^U & H_{33} \\ \end{array}\right] \nonumber \\
\lambda^D &=& \left[\begin{array}{ccc} 0 & H_{12} x_{12}^D & 0\\
H_{21}^D & H_{22} x_{22}^D & 0 \\
0 & H_{32} x_{32}^D & H_{33} \\ \end{array}\right] \nonumber \\
\lambda^E &=& \left[\begin{array}{ccc} 0 & H_{12} x_{12}^E & 0\\
H_{21}^E & H_{22} x_{22}^E & 0 \\
0 & H_{32} x_{32}^E & H_{33} \\ \end{array}\right], \label{mxyuks}
\end{eqnarray}
where we have defined
\begin{eqnarray}
H_{21}^U e^{i \phi_{21}^U} &\equiv& H_{21}
x_{21}^U e^{i\phi_{21}}
+ \tilde{H}_{21} \tilde{x}_{21}^U e^{i\tilde{\phi}_{21}}  \nonumber \\
H_{21}^{D,E} &\equiv& H_{21}
x_{21}^{D,E} e^{i\phi_{21}}
+ \tilde{H}_{21} \tilde{x}_{21}^{D,E} e^{i\tilde{\phi}_{21}} 
\end{eqnarray}
We may now remove ${\phi_{22}}'$ by phase transformations upon
$\bar{F}_{1,2,3}$ but $\phi_{21}^U$ may only be removed by a phase
redefinition of $F_{1,2,3}$, which would alter the prediction of the
CKM matrix $V_{CKM}$. Thus, $\phi_{21}^U$ is a physical phase, that is
it cannot be completely removed by phase rotations upon the fields.
Once the operators $O_{21},\tilde{O}_{21}, 
O_{12}$ have been chosen, the Yukawa
matrices at  $M_{GUT}$ including the phase in the CKM matrix are
therefore
identified with $H_{ij}, {H'}_{22},\phi_{21}^U$.

\section{Numerical Analysis of Masses and Mixing Angles 
from Non-Symmetric  Textures \label{numerical}}

In this section we discuss the numerical procedure used to
analyse the non-symmetric cases introduced in the previous section. 
We shall perform an analysis on the new approach based on
$n=1$ operators only, and also re-analyse and up-date the original scheme
of ref.\cite{422} for comparison.

The basic idea is to do a global fit of each considered Ansatz to
$m_e,m_\mu$, 
$m_u$,
$m_c$,
$m_t$,
$m_d$,
$m_s$,
$m_b$,
$\alpha_S(M_Z)$,
$|V_{ub}|$, $|V_{cb}|$ and
$|V_{us}|$ using $m_\tau$ as a constraint. We use the approximation that the
whole SUSY spectrum of the MSSM lies at $M_{SUSY}=m_t$
and that the MSSM remains a valid effective theory until the scale
$M_{GUT}=10^{16}$ GeV. Not wishing to include neutrino masses in this
analysis, we simply set the right-handed Majorana neutrino mass of
each family to be $10^{16}$ GeV so that the neutrinos are
approximately massless and hence their masses do not affect the RGEs
below $M_{GUT}$. Recall the parameters introduced in Eq.\ref{mxyuks}:
$\phi_{21}^U \equiv \phi$, $H^U_{21}\equiv {H_{21}}'$, 
$H^D_{21}\equiv H_{21}$, 
${H_{22}}', H_{22}$, $H_{12}$,
$H_{32}, H_{33}$. The values of these 8 parameters
plus $\alpha_S$ at the GUT scale are determined by the fit.

The matrices $\lambda^I$ are diagonalised numerically and 
$|V_{ub}(M_{GUT})|, |V_{us}(M_{GUT})|$ are determined by 
\begin{equation}
V_{CKM} = {V_U}_L {V_D}_L^\dagger,
\end{equation}
where ${V_U}_L, {V_D}_L$ are the matrices that act upon the
$(u,c,t)_L$ and $(d,s,b)_L$ column
vectors respectively to transform from the weak eigenstates to the mass
eigenstates of the quarks. We use the boundary conditions
$\alpha_1(M_{GUT})=\alpha_2(M_{GUT})=0.708$, motivated by previous
analyses based on gauge unification in SUSY GUT models \cite{GUTun}.
$\lambda_{u,c,t,d,s,b,e,\mu,\tau}$, $|V_{us}|$ and $|V_{ub}|$ are
then run\footnote{All
renormalisation running in this paper is one loop and in the
$\overline{\mbox{MS}}$ scheme. The relevant renormalisation group
equations (RGEs)
are listed in ref.\cite{422}.} from $M_{GUT}$ to 170 GeV$\approx m_t$
using the
RGEs for the MSSM\@. 
Below $M_{GUT}$ the
effective field theory of the Standard Model allows the couplings in
the different charge sectors to split and run differently. 
The $\lambda_i$ are then evolved to their
empirically derived running masses using 3 loop QCD$\otimes$1 loop
QED \cite{422}. 
$m_\tau^{e}$ and $\lambda_\tau^{p}(m_\tau)$ then\footnote{The
superscript $e$ upon masses, mixing angles or diagonal Yukawa
couplings denotes an empirically derived value, whereas the
superscript $p$ denotes the prediction of the model for the 
particular fit
parameters being tested.}
fix $\tan \beta$ through
the relation~\cite{Andersonetal}
\begin{equation}
\cos \beta = \frac{\sqrt{2} m_{\tau}^e (m_\tau)}{v \lambda_\tau^p
(\lambda_\tau)},
\label{cosbeta}
\end{equation}
where $v=246.22$ GeV is the VEV of the Standard Model Higgs.
Predictions of the other fermion masses then come from
\begin{eqnarray}
m_{c,t}^p &\approx& \lambda_{c,t}^p (m_{c,t}) \frac{v \sin
\beta}{\sqrt{2}},\nonumber \\
m_{d,s,b}^p &\approx& \lambda_{d,s,b} (m_{1,1,b}^e) \frac{v \cos
\beta}{\sqrt{2}} \nonumber \\
m_{e,\mu}^p &\approx& \lambda_{e,\mu} (m_{1,\mu}^e) \frac{v \cos
\beta}{\sqrt{2}}, \label{masspred}
\end{eqnarray}
where $m_1 \equiv$ 1 GeV.
There are twelve data points and nine parameters so we
have three degrees of freedom (dof). The parameters are all varied until the
global $\chi^2 / \mbox{dof}$ is minimised. The data used (with
1$\sigma$ errors quoted) is \cite{databook}
\begin{eqnarray}
m_e &=& 0.510999 \mbox{~MeV} \nonumber \\
m_\mu &=& 105.658   \mbox{~MeV} \nonumber \\
m_\tau &=& 1.7771 \mbox{~GeV} \nonumber \\
m_c &=& 1.3 \pm 0.3 \mbox{~GeV} \nonumber \\
m_t^{phys} &=& 180 \pm 12 \mbox{~GeV} \nonumber \\
m_d &=& 10 \pm 5  \mbox{~MeV} \nonumber \\
m_s &=& 200 \pm 100  \mbox{~MeV} \nonumber \\
m_b &=& 4.25 \pm 0.1  \mbox{~GeV} \nonumber \\
|V_{ub}| &=& (3.50 \pm 0.91) 10^{-3} \nonumber \\
|V_{us}| &=& 0.2215 \pm 0.0030 \nonumber \\
\alpha_S(M_Z) &=& 0.117 \pm 0.005 \label{data}
\end{eqnarray}
$|V_{cb}|$ is fixed by $H_{32}$ which does not influence the other
predictions to a good approximation and so $|V_{cb}|$ and $H_{32}$
effectively decouple from the fit. We merely note that in all cases,
to predict the measured value of $|V_{cb}|$, $H_{32} \sim 0.03$.
Note that no errors are quoted upon the lepton masses because $m_\tau$
is used as a constraint on the data and because $m_e, m_\mu$ were
required to be satisfied to 0.1\% by the fit. 
In this way we merely use the lepton masses as 3
constraints, using up 3 dof. We did not perform the
fit with smaller empirical errors on the lepton masses
because of the numerical
roundoff and minimisation errors associated with high $\chi^2$ values
generated by them. Also, 0.1\% is a possible estimate of higher loop
radiative corrections
involved in the predictions. Note that no other theoretical errors
were taken into account in the fit. The largest ones may occur in
derivations of $m_b$ due to the large $\lambda_b$ coupling~\cite{mb}
and the non-perturbative effects of QCD near 1 GeV. It is not clear
how to estimate these errors since the error on $m_b$ depends upon
soft parameters which depend on the SUSY breaking mechanism in a very
model dependent way and non-perturbative QCD is an unsolved problem.
The correlations between the empirical estimations of the current
quark masses are also not included.
A potentially large error could occur if the ansatze considered are
not exact but are subject to corrections by higher dimension
operators. We discuss this point further in
section~\ref{familysymmetry}.

The results obtained from this analysis are given in
Table~\ref{tab:res1}. 
\begin{table}
\begin{center}
\begin{tabular}{|c|c|c|c|c|c|} 
\hline
Model & 1 & 2 & 3 & 4 & 5 \\
\hline
 $O_{12}$ & $O^M$ & $O^W$ & $O^R$ & $O^R$ & $O^R$ \\
 $O_{21}+\tilde{O}_{21}$ & $O^M+O^A$ &$O^G+O^H$ &$O^M+O^A$ &$O^G+O^H$
&$O^R+O^S$ \\
\hline
 $H_{22}/10^{-2}$ & 2.88 & 2.64 & 2.69 & 2.67 & 6.15 \\
 $H_{12}/10^{-3}$ & 2.81 & 4.41 & 2.13 & 0.70 & 1.21 \\
 $H_{21}/10^{-3}$ & 1.30 & 5.97 & 1.76 & 4.33 & 1.91 \\
 $\cos \phi$ & 0.87 & 1.00 & 0.20 & 1.00 & 0.61 \\
 $H_{33}$ & 1.18 & 1.05 & 1.05 & 1.07 & 4.6 \\
 ${H_{22}}'/10^{-3}$ & 1.91 & 1.87 & 1.87 & 1.87 & 2.87 \\
 ${H_{21}}'/10^{-3}$ & 1.94 & 1.62 & 1.63 & 1.66 & 0.76 \\
\hline
 $\alpha_S(M_Z)$ & 0.119& 0.118 & 0.118 & 0.118 & 0.126 \\
$m_d$/MeV & 6.25 & 1.03 & 8.07 & 4.14 & 11.9 \\   
$m_s$/MeV & 158 & 150 & 154 & 152 & 228 \\
$m_c$/GeV & 1.30 & 1.30 & 1.30 & 1.30 & 1.30 \\
$m_b$/GeV & 4.24 & 4.25 & 4.25 & 4.25 & 4.13 \\
$m_t^{phys}$/GeV & 182 & 180 & 180 & 180 & 192 \\
$|V_{us}|$ & 0.2211 & 0.2215 & 0.2215 & 0.2215 & 0.2215 \\
$|V_{ub}|/10^{-3}$ & 3.71 & 3.51 & 3.50 & 3.52 & 3.50 \\
$\tan \beta$ & 59.5 & 58.3 & 58.3 & 58.5 & 65.7 \\ \hline
$\chi^2$/dof & 0.34 & 1.16 & 0.13 & 0.55 & 1.84 \\
\hline
\end{tabular}
\end{center}
\caption{{\small Results of best-fit analysis on models with $n=1$ operators
only. Note
that the input parameters $H_{ij}, {H_{ij}}', \cos \phi$ shown are
evaluated at the scale $M_{GUT}$. All of the mass predictions shown are
running masses, apart from the pole mass of the top
quark\protect\footnote{To one loop.},
$m_t^{phys}\equiv m_t(1 + \frac{4 \alpha_S(m_t)}{3
\pi})$. The CKM matrix element
predictions are at $M_Z$.}}
\label{tab:res1}
\end{table}
Out of 16 possible models that fit the texture required by
Eqs.\ref{cleb},\ref{dom}, 11 models fit the data with
$\chi^2$/dof$<3$. Out of these 11 models, 5 fit the data with
$\chi^2$/dof$<2$ and these are displayed in Table~\ref{tab:res1}.
The operators listed as $O_{12},O_{21},\tilde{O}_{21}$ describe the
structure of the
models and the entries $H_{22}, H_{12}, H_{21}, \cos \phi, H_{33},
{H_{22}}', {H_{21}}'$ are the GUT scale input parameters of the best
fit values of the model. The estimated 1$\sigma$ deviation in
$\alpha_S(M_Z)$ from the fits is $\pm 0.003$ and the other parameters
are constrained to better than 1\% apart from $\cos \phi$, whose
1$\sigma$ fit 
errors often cover the whole possible range.
Out of the predictions shown in Table~\ref{tab:res1}, $m_d$
discriminates between the models the widest. $\alpha_S(M_Z)$ takes
roughly central values, apart from model 5 for which the best fit is
outside the 1$\sigma$ errors quoted in Eq.\ref{data} on
$\alpha_S(M_Z)$. $m_s, |V_{ub}|$ are
within 1$\sigma$ of the data point and $m_c,|V_{us}|$ are
approximately on the central value for all 5 models. Models 3,1 and 4
are very satisfactory fits to the data with $\chi^2$/dof$<1$.
We conclude that the $\chi^2$ test has some discriminatory power in
this case since if all of the models were equally good, we would
statistically expect to have 11 models with $\chi^2$/dof$<1$, 3 models
with $\chi^2$/dof$=1-2$ and 2 models with $\chi^2=2-3$ out of the 16
tested.

We now briefly return to the original models with upper blocks
given by $B_{1-8}$ in Eqs.\ref{suclighti}-\ref{suclightii} \cite{422}. 
After again isolating the only physical phase to
$\lambda_{21}^U$, a numerical fit analogous to the above was
performed using the same data in Eq.\ref{data}. The main difference in
the fit with these models is that there are now 4 degrees of freedom
in the fit (since there is one less parameter). All eight models in
question fit the data with $\chi^2 < 2$ and these are displayed in
Table~\ref{tab:res2}.
\begin{table}
\begin{center}
\begin{tabular}{|c|c|c|c|c|c|c|c|c|} 
\hline
Model & $B_1$ & $B_2$ & $B_3$ & $B_4$ & $B_5$ & $B_6$ & $B_7$ & $B_8$\\
\hline
$\alpha_S(M_Z)$ & 0.123& 0.123 & 0.123 & 0.124 & 0.123 & 0.124 & 0.125
& 0.124 \\
$m_d$/MeV & 7.58 & 9.12 & 4.64 & 6.18 & 7.49 & 3.63 & 3.53 & 4.53 \\
$m_s$/MeV & 215 & 240 & 179 & 210 & 217 & 179 & 200 & 187 \\
$m_c$/GeV & 1.29 & 1.38 & 1.35 & 1.16 & 1.29 & 1.32 & 0.86 & 1.31 \\
$m_b$/GeV & 4.19 & 4.17 & 4.19 & 4.19 & 4.19 & 4.18 & 4.20 & 4.19 \\
$m_t^{phys}$/GeV & 188 & 189 & 189 & 189 & 188 & 189 & 190 & 189 \\
$|V_{us}|$ & 0.2212 & 0.2213 & 0.2214 & 0.2212 & 0.2212 & 0.2215 &
0.2212 & 0.2214\\
$|V_{ub}|/10^{-3}$ & 4.52 & 4.37 & 4.05 & 4.22 & 4.56 & 3.74 & 3.85 &
3.98 \\
$\tan \beta$ & 63.2 & 63.6 & 63.4 & 63.7 & 63.2 & 63.8 & 64.3 & 63.6
\\
\hline
$\chi^2$/dof & 0.95 & 0.96 & 1.00 & 1.05 & 0.97 & 1.16 & 1.87 & 1.04 \\
\hline
\end{tabular}
\end{center}
\caption{{\small Predictions of best-fit analysis on models from
ref.~\protect\cite{422} with $n=2$
operators included. 
All of the mass predictions shown are
running masses, apart from the pole mass of the top
quark. The CKM matrix element
predictions are at $M_Z$.}}
\label{tab:res2}
\end{table}
We do not display the best fit input parameters because they are
largely irrelevant for the discussion here. 1$\sigma$ fit deviations
of $\alpha_S(M_Z)$ are again 0.003 for $B_{1-8}$. Note that whereas these
models are able to fit $|V_{us}|, m_s, m_d, m_b, m_c$ fairly well, their
predictions of $\alpha_S(M_Z)$ are high and outside the 1$\sigma$
empirical error bounds. $|V_{ub}|$ is naturally high in these models
(as found in ref.\cite{422}) and this forces $\alpha_S(M_Z)$ to be
large, where $|V_{ub}|$ may decrease somewhat.
To fit $m_b$ with a high $\alpha_S(M_Z)$
requires a large $H_{33}$ element and this is roughly speaking why
$m_t^{phys}$ is predicted to be quite high. In each model the high
value of $\alpha_S (M_Z)$ required is the dominant source of $\chi^2$
apart from $B_7$, where $m_c$ is low.

In comparison to the new scheme with $n=1$ operators only, the old
scheme with $n=2$ operators fits the data
pretty well, although not quite as well as models 1,3,4.
The old scheme also has one more prediction than the new one.
However, the preferred models are the ones incorporating the $U(1)_X$
symmetry since they go deeper into the reasons for the zeroes and
hierarchies in the Yukawa matrices.

\section{\mbox{\boldmath{$U(1)_X$}} Family Symmetry
in the $SU(4)\times O(4)$ Model \label{familysymmetry}}

In our discussion of the symmetric textures, we assumed
that we could obtain the same structure as IR\@. Of course,
as we have already mentioned, the case we are examining is
different in two aspects: (a) the fermion mass matrices of the different
charge sectors 
have the same origin, and thus the same expansion
parameter and (b) all differences between these sectors
arise from Clebsch factors.
As a starting point, we will therefore
briefly repeat the IR analysis for {\bf symmetric mass
matrices} in our framework;
we then go on to consider the non-symmetric case, with the
goal of being able to reproduce the numerical values
(at least to an order of magnitude) of the successful ansatze
given in the previous section.

The structure of the mass
matrices is determined by a family symmetry, $U(1)_{X}$,
with the charge assignment of the various states given
in Table~\ref{table:1}. 
\begin{table}[h]
\centering
\begin{tabular}{|c |cccccccccc|}\hline
   &$ Q_i$ & $u^c_i$ &$ d^c_i$ &$ L_i$ & $e^c_i$ & $\nu^c_i$ &
$h_1$ &
$ h_2$ & $H$ & $\bar{H}$   \\
\hline
  $U(1)_{X}$ & $\alpha _i$ & $\alpha _i$ & $\alpha _i$  & $\alpha_i$
& $\alpha_i$ & $\alpha_i $ & $-2\alpha _3$ &  $-2\alpha _3$
& x & -x 
\\
\hline
\end{tabular}
\caption{{\small $U(1)_{X}$ charges assuming symmetric textures.}
\label{table:1}}
\end{table}
The need to preserve $SU(2)_L$
invariance requires left-handed up and down quarks (leptons)
to have the same charge. This, plus the additional
requirement of symmetric
matrices, indicates that all quarks (leptons) of the same i-th
generation transform with the same charge $\alpha _i$.  
Finally, lepton-quark unification under 
$SU(4) \otimes SU(2)_L \otimes SU(2)_R$  indicates that
quarks and leptons of the same family have the same charge
(this is a different feature as compared to IR, where quarks
and leptons of the two lower generations have 
different charges under the flavour symmetry).
The full anomaly free Abelian group involves an additional family
independent component, $U(1)_{FI}$, and with this freedom
$U(1)_{X}$ is made traceless without any loss of 
generality\footnote{Since we assume that the 33 operator is 
renormalisable, the relaxation of the
tracelessness condition does not change the
charge matrix since any additional FI charges can always be absorbed
into the Higgs $h_i$ charges.}.
Thus we set $\alpha_1=-(\alpha_2+\alpha_3)$. Here we consider
the simplest case where the combination
$H\bar{H}$ is taken to have zero charge. This is consistent
with our requirement that it plays no
role in the mass hierarchies, other than leading to
a common factor $\delta$ for all non-renormalisable entries.

 If the light Higgs $h_{2}$, $h_{1}$, responsible for the up and
down quark masses respectively, arise from the same bidoublet
$h=(1,2,2)$, then they have the same $U(1)_X$ charge so that only 
the 33 renormalisable Yukawa coupling to $h_{2}$, $h_{1}$ is
allowed, and only the 33 element of the associated mass matrix
will be non-zero.  The remaining entries are generated when the
$U(1)_X$ symmetry is broken. This breaking is taken to be spontaneous,
via Standard Model singlet fields, which can be either {\bf chiral}
or {\bf vector} ones; in the latter case, which is the one
studied in IR, two fields
$\theta,\; \bar{\theta}$, with $U(1)_X$ charge -1, +1
respectively and 
equal VEVs are introduced. When these
fields get a VEV, the mass matrix acquires its structure. For
example, the 32 - entry in the up quark mass matrix appears at
$O(\epsilon )$ because  U(1) charge conservation only allows 
the term $c^c t h_2({\theta}/M_2)^{\alpha_2-\alpha_3}$ for
$\alpha_2>\alpha_3$, or $c^ct
h_2(\bar{\theta} /M_2)^{\alpha_3-\alpha_2},$ for $\alpha_3>\alpha_2$.
Here
$\epsilon=(<\theta>/M_2)^{\mid\alpha_2-\alpha_3\mid}$
where $M_2$ is the unification mass
scale  which governs the higher dimension operators. In IR,
a different scale, $M_{1}$, is expected
for the down quark and lepton mass matrices.

In our case however,  all charge and mass
matrices have the same structure under the
$U(1)_{X}$ symmetry, since all known fermions are
accommodated in the same multiplets of the gauge group.
The charge matrix is of the form
\begin{eqnarray}
\left(
\begin{array}{ccc}
-2 \alpha_2 - 4 \alpha_3 & 
-3 \alpha_3 & -\alpha_2 - 2\alpha_3 \\
-3 \alpha_3 & 2(\alpha_2-\alpha_3) 
& \alpha_2 - \alpha_3 \\
-\alpha_2 -2 \alpha_3 & \alpha_2 - \alpha_3 & 0
\end{array}
\right)
\end{eqnarray}
Then, including the
common factor $\delta$ which multiplies
all non-renormalisable entries,
the following pattern of masses is obtained
(for vector-like singlets):
\begin{eqnarray}
\lambda^{u,d,\ell}\approx 
\left(
\begin{array}{ccc}
\delta \epsilon^{\mid 2+6a \mid } &
\delta \epsilon^{\mid 3a \mid } &
\delta \epsilon^{\mid 1+3a\mid }
\\
\delta \epsilon^{\mid 3a \mid } &
\delta \epsilon^{ 2 } &
\delta \epsilon \\
\delta \epsilon^{\mid 1+3a \mid } &
\delta \epsilon & 1
\end{array}
\right)
, \; \; \; \;
\label{eq:massu}
\end{eqnarray}
where\footnote{In this simplest (and more
predictive)
realisation, $h_b\approx h_{t}$
therefore we are in the large $\tan\beta$ regime of the
parameter space of the MSSM\@.}
$a=\alpha_3/(\alpha_2-\alpha_3)$. 
We emphasise that the entries in Eq.\ref{eq:massu}
describe the magnitudes of the dominant
operators, and do not take the Clebsch zeroes of the different charge
sectors into account.
Note the existence of 
a single expansion parameter,
for all three matrices. Another interesting point is
that a unique charge combination $a$ appears
in the exponents of all matrices, as a result
of quark-lepton unification. 
Actually, unlike what appears here, in most schemes the
lepton mass matrix is described in
the generic case by two parameters. For $a=1$, one generates the structure in
Eq.~\ref{deltaIR2}
for the unified fermion mass matrices.

Before passing to the non-symmetric case, let us 
make a few comments on the possibility of
having chiral or vector singlets, as well as
on the charge of the Higgs fields.
Suppose first that $\theta$ is a chiral field.
{}From the form of the charge matrix, we observe that
if the 22 and 23 entries have a positive
charge, $\alpha_3$ is negative (for all these entries to
be non-vanishing at the same time).
Moreover the hierarchy 1:3 between the 23
and 12 elements indicates
that $\alpha_2$ would have to be zero in the chiral case,
and thus the 13 element would tend to be larger than
desired.
We can say therefore that in the symmetric case 
with vector fields generates the mass hierarchies in
a more natural way.

Concerning the $h_1$, $h_2$ higgses, 
there are two kinds originating from free fermionic string models:
those coming from
Neveu-Schwarz sector 
which in general have integer (including zero)  $U(1)_X$ charges, 
and those arising from twisted sectors, which
usually carry fractional $U(1)_X$ charges. 
Which of these cases acquire VEVs, is 
decided from the phenomenological analysis.
For example, 
to obtain the structure of
Eq.\ref{deltaIR2} we see that the charges of $h_{1,2}$ may
not be zero, since in such a case the 12 element which
is proportional to the Higgs charge would be unacceptably large.
For the non-symmetric case of course this feature does not
necessarily hold.
Finally, the $H$, $\bar{H}$ fields (the $SU(4)$ higgses) tend to be
non-singlets under extra $U(1)_X$ symmetries. 
We now proceed to discuss the {\bf non-symmetric case}, which
in the framework of $U(1)_X$ symmetries has been
extensively studied in \cite{blr}. Here, we will examine
what constraints one may put on the various
possibilities for non-symmetric textures, in the model under study.

The charge assignment for this case appears 
in Table~\ref{table:2}.
\begin{table}[h]
\centering
\begin{tabular}{|c |cccccccccc|}\hline
   &$ Q_i$ & $u^c_i$ &$ d^c_i$ &$ L_i$ & $e^c_i$ & $\nu^c_i$ &
$h_1$ &
$ h_2$ & $H$ & $\bar{H}$   \\
\hline
 $U(1)_{X}$ & $\beta _i$ & $\alpha _i$ & $\alpha _i$  & $\beta_i$
& $\alpha_i$ & $\alpha_i $ & $-\beta _3-\alpha_3 $ &  
$-\beta _3- \alpha_3$ & x & -x 
\\
\hline
\end{tabular}
\caption{{\small $U(1)_{X}$ charges for non-symmetric textures.}
\label{table:2}}
\end{table}
Fields that belong to the same representation
of $SU(4) \otimes SU(2)_L \otimes SU(2)_R$
are taken to have the same charge.
Again, it is clear that all fermion
mass matrices will have the same structure.
With this charge assignment we may proceed as in the
symmetric case, and calculate the possible mass matrices that may
arise. The charge matrix is now
\begin{eqnarray}
\left(
\begin{array}{ccc}
-\alpha_2 -2 \alpha_3 - \beta_2 -2 \beta_3 &
\alpha_2 - \alpha_3 - \beta_2 - 2 \beta_3 &
-\beta_2 - 2 \beta_3 \\
-\alpha_2 - 2 \alpha_3 + \beta_2 - \beta_3 &
\alpha_2 - \alpha_3 + \beta_2 - \beta_3 &
\beta_2 - \beta_3  \\
-\alpha_2 - 2 \alpha_3 &
\alpha_2 - \alpha_3 & 0 
\end{array}
\right)
\end{eqnarray}
We now want to find which charge assignments may generate
a mass matrix as close as possible to the form in
Eq.\ref{deltanonsym},
keeping in mind that
there is no reason to restrict ourselves to non-symmetric textures
with a zero in the 13 and 31 position.

In what follows, we will check whether it is possible to
generate the hierarchies in the effective low energy Yukawa couplings required
by our ansatze and the data. The required couplings are detailed in
Table~\ref{tab:res1}. Initially, we determine if we can obtain the correct
structure by chiral singlet fields.
We assume for a starting point that for the 32 entry
we have 
$\alpha_2 - \alpha_3 > 0$ (without a loss of generality
since we can always choose the sign of one entry in the
charge matrix). The 23 entry has to be small (it is assumed to be zero in the
ansatze in Eq.\ref{dom}), indicating
that (a) either $\beta_2 - \beta_3 < 0$ or
(b) $\beta_2 - \beta_3$ is positive and
large ($\geq 2$). Case (b) is excluded, since it would indicate that
the 22 charge, which is always the sum
of the 23 and 32 charges, would be
unacceptably large as well (which implies that $H_{22} < H_{32}$, in
contradiction
to the fits in Table~\ref{tab:res1}). What about case (a)?
A negative number must not dominate the 22 entry
in the chiral case, thus 
$|\beta_2 - \beta_3|$ would have to be smaller than
$|\alpha_2 - \alpha_3|$. This clearly contradicts the
required hierarchy between the 22 and 32 elements and so the required
couplings can not be naturally described by a model with only a chiral
U(1)$_X$ Higgs $\theta$.

For this reason we are going to look for solutions in the case
of vector singlets, where it is the absolute value of the
charges that matters. Here, the important difference from the
previous case
is that a solution with a 
small and positive $\alpha_2 - \alpha_3$
and a large negative $\beta_2 - \beta_3$
is allowed. The 23 and 32 elements
have the correct hierarchy, while the 22
element can also be sufficiently 
small, as a result of a cancellation
between terms of opposite sign, with the negative
contribution being dominant.
What can we say about the rest of the structure and how restrictive
should we be when looking for solutions?
We could allow for a 
small asymmetry between the 12 and 21 entries.
Actually, $\lambda_{12}^D$ can be slightly larger
than $\lambda_{21}^D$. This, combined with 
the fact that there are unknown coefficients of
order unity 
indicates that we can have an asymmetry of order $\epsilon$
between the 12 and 21 entries. We will discuss solutions
with such an asymmetry, even in the case that
$\lambda_{12}^D < \lambda_{21}^D$, due to this coefficient
ambiguity as well as the ambiguity in the experimental
value of the up and down quark.
We also need not drop solutions with a large
13 or 31 entry, if they are compatible with the numerics.

On this basis, we have looked for solutions in the
following way: for the charges of the elements
12-21-22-32 we made all possible charge
assignments (such that lead 
to a maximum $4^{th}$ power in terms of the
expansion parameter for the resulting
mass matrices, for the 12
and 21 entries). This
fixes all charges $\alpha_2$,
$\alpha_3$, $\beta_2$,
$\beta_3$ each time. We then looked at what 
the charges of the other entries are
and whether the generated
hierarchies are consistent with the phenomenology.

The restrictions we require in order to identify a 
viable solution, are
(besides of course that the only renormalisable term
is in the 33 position)
\begin{eqnarray}
     |\mbox{charge}(11)| & > & |\mbox{charge}(12)| \nonumber \\
     |\mbox{charge}(11)| & > & |\mbox{charge}(21)| \nonumber \\
     |\mbox{charge}(21)| & > & |\mbox{charge}(22)| \nonumber \\
     |\mbox{charge}(12)| & > & |\mbox{charge}(22)| \nonumber \\
     |\mbox{charge}(13)| & > & |\mbox{charge}(22)| \nonumber \\
     |\mbox{charge}(31)| & > & |\mbox{charge}(22)| \nonumber \\
     |\mbox{charge}(32)| & \leq & |\mbox{charge}(22)| 
     O ( \epsilon ) 
                                       \nonumber \\
     |\mbox{charge}(12)| & \approx & |\mbox{charge}(21)|   O ( \epsilon ) 
                                      \nonumber \\
     |\mbox{charge}(23)| & > & |\mbox{charge}(22)| 
\end{eqnarray}
Then, we end up with the following possibilities:\\
{\bf Case 1:} 
\begin{equation}
\alpha_2 = -2/3, \; \;  
\alpha_3 = -5/3, \; \;   
\beta_2 = -2, \; \;  
\beta_3 = 0, \; \;  
Y_{u,d,\ell} = 
\left(\begin{array}{ccc}
\delta {\epsilon}^6 & \delta {\epsilon}^3 & \delta {\epsilon}^2 \\
\delta {\epsilon}^2 & \delta {\epsilon} & \delta {\epsilon}^2 \\
\delta {\epsilon}^4 & \delta {\epsilon} & 1 \\ 
\end{array}\right) 
\end{equation}
{\bf Case 2:} 
\begin{equation}
\alpha_2 = -1, \; \;  
\alpha_3 = -2, \; \;   
\beta_2 = -2, \; \;  
\beta_3 = 0, \; \;  
Y_{u,d,\ell} = 
\left(\begin{array}{ccc}
\delta {\epsilon}^7 & \delta {\epsilon}^3 & \delta {\epsilon}^2 \\
\delta {\epsilon}^3 & \delta {\epsilon} & \delta {\epsilon}^2 \\
\delta {\epsilon}^5 & \delta {\epsilon} & 1 \\ 
\end{array}\right) 
\end{equation}
{\bf Case 3:} 
\begin{equation}
\alpha_2 = -4/3, \; \;  
\alpha_3 = -7/3, \; \;   
\beta_2 = -2, \; \;  
\beta_3 = 0, \; \;  
Y_{u,d,\ell} = 
\left(\begin{array}{ccc}
\delta {\epsilon}^8 & \delta {\epsilon}^3 & \delta {\epsilon}^2 \\
\delta {\epsilon}^4 & \delta {\epsilon} & \delta {\epsilon}^2 \\
\delta {\epsilon}^6 & \delta {\epsilon} & 1 \\ 
\end{array}\right) 
\end{equation}
{\bf Case 4:} 
\begin{equation}
\alpha_2 = -4/3, \; \;  
\alpha_3 = -1/3, \; \;   
\beta_2 = 0, \; \;  
\beta_3 = -2, \; \;  
Y_{u,d,\ell} = 
\left(\begin{array}{ccc}
\delta {\epsilon}^6 & \delta {\epsilon}^3 & \delta {\epsilon}^4 \\
\delta {\epsilon}^4 & \delta {\epsilon} & \delta {\epsilon}^2 \\
\delta {\epsilon}^2 & \delta {\epsilon} & 1 \\ 
\end{array}\right) 
\end{equation}
{\bf Case 5:} 
\begin{equation}
\alpha_2 = -4/3, \; \;  
\alpha_3 = -7/3, \; \;   
\beta_2 = -3, \; \;  
\beta_3 = 0, \; \;  
Y_{u,d,\ell} = 
\left(\begin{array}{ccc}
\delta {\epsilon}^9 & \delta {\epsilon}^4 & \delta {\epsilon}^3 \\
\delta {\epsilon}^3 & \delta {\epsilon}^2 & \delta {\epsilon}^3 \\
\delta {\epsilon}^6 & \delta {\epsilon} & 1 \\ 
\end{array}\right) 
\end{equation}
{\bf Case 6:} 
\begin{equation}
\alpha_2 = -1, \; \;  
\alpha_3 = -2, \; \;   
\beta_2 = -7/3, \; \;  
\beta_3 = -1/3, \; \;  
Y_{u,d,\ell} = 
\left(\begin{array}{ccc}
\delta {\epsilon}^8 & \delta {\epsilon}^4 & \delta {\epsilon}^3 \\
\delta {\epsilon}^3 & \delta {\epsilon} & \delta {\epsilon}^2 \\
\delta {\epsilon}^5 & \delta {\epsilon} & 1 \\ 
\end{array}\right) 
\end{equation}
{\bf Case 7:} 
\begin{equation}
\alpha_2 = -5/3, \; \;  
\alpha_3 = -8/3, \; \;   
\beta_2 = -3, \; \;  
\beta_3 = 0, \; \;  
Y_{u,d,\ell} = 
\left(\begin{array}{ccc}
\delta {\epsilon}^{10} & \delta {\epsilon}^4 & \delta {\epsilon}^3 \\
\delta {\epsilon}^4 & \delta {\epsilon}^2 & \delta {\epsilon}^3 \\
\delta {\epsilon}^7 & \delta {\epsilon} & 1 \\ 
\end{array}\right) 
\end{equation}
{\bf Case 8:} 
\begin{equation}
\alpha_2 = -4/3, \; \;  
\alpha_3 = -7/3, \; \;   
\beta_2 = -7/3, \; \;  
\beta_3 = -1/3, \; \;  
Y_{u,d,\ell} = 
\left(\begin{array}{ccc}
\delta {\epsilon}^9 & \delta {\epsilon}^4 & \delta {\epsilon}^3 \\
\delta {\epsilon}^4 & \delta {\epsilon} & \delta {\epsilon}^2 \\
\delta {\epsilon}^6 & \delta {\epsilon} & 1 \\ 
\end{array}\right) 
\end{equation}
{\bf Case 9:} 
\begin{equation}
\alpha_2 = -4/3, \; \;  
\alpha_3 = -1/3, \; \;   
\beta_2 = -1/3, \; \;  
\beta_3 = -7/3, \; \;  
Y_{u,d,\ell} = 
\left(\begin{array}{ccc}
\delta {\epsilon}^7 & \delta {\epsilon}^4 & \delta {\epsilon}^5 \\
\delta {\epsilon}^4 & \delta {\epsilon} & \delta {\epsilon}^2 \\
\delta {\epsilon}^2 & \delta {\epsilon} & 1 \\ 
\end{array}\right) 
\end{equation}
Let us also list for completeness a few cases
with a larger splitting between the 21 and 
12 entries (up to $O(\epsilon^{2})$ ):\\
{\bf Case 10:} 
\begin{equation}
\alpha_2 = -4/3, \; \;  
\alpha_3 = -1/3, \; \;   
\beta_2 = 1/3, \; \;  
\beta_3 = -5/3, \; \;  
Y_{u,d,\ell} = 
\left(\begin{array}{ccc}
\delta {\epsilon}^5 & \delta {\epsilon}^2 & \delta {\epsilon}^3 \\
\delta {\epsilon}^4 & \delta {\epsilon} & \delta {\epsilon}^2 \\
\delta {\epsilon}^2 & \delta {\epsilon} & 1 \\ 
\end{array}\right) 
\end{equation}
{\bf Case 11:} 
\begin{equation}
\alpha_2 = -2/3, \; \;  
\alpha_3 = -5/3, \; \;   
\beta_2 = -7/3, \; \;  
\beta_3 = -1/3, \; \;  
Y_{u,d,\ell} = 
\left(\begin{array}{ccc}
\delta {\epsilon}^7 & \delta {\epsilon}^4 & \delta {\epsilon}^3 \\
\delta {\epsilon}^2 & \delta {\epsilon} & \delta {\epsilon}^2 \\
\delta {\epsilon}^4 & \delta {\epsilon} & 1 \\ 
\end{array}\right) 
\end{equation}

Of course, here we also have the cases with the opposite
charge assignment\footnote{The presence of fractional charges implies
the existence
of residual discrete symmetries after the breaking of the
abelian symmetry.}. Among the various choices, we
see that
\begin{itemize}
\item{The charge of the Higgs fields $h_{1,2}$ is always
different from zero.}
\item{There are cases where the 13 and 31 elements are
large.}
\end{itemize}
We may now examine the results of Table~\ref{tab:res1} in the context
of the U(1)$_X$ symmetry discussion above. We take all models that fit
the data with $\chi^2$/dof$<1$, i.e.\ models 1,3,4. We define in each of these
models, $H^{emp}_{ij}$ as being the dimensionless and dominant
effective coupling
constants in the SU(4)$\otimes$SU(2)$_L\otimes$SU(2)$_R$ unified
Yukawa matrix for the best fit parameters. 

Then, model 1 has
\begin{equation}
H_{ij}^{emp} \sim \left[\begin{array}{ccc}
0 & 0.003 & 0 \\
0.001 & 0.03 & 0\\
0 & 0.03 & 1\\ \end{array}\right]. \label{mod1}
\end{equation}
We see that case 1 above does not fit this pattern very well if all
dimensionless couplings are $\sim O(1)$ because
in case 1, $H_{21}$ is suppressed in comparison to $H_{12}$. Cases 4,9
do not possess approximate texture zeroes in the 31 position and this
would affect $|V_{ub}|$ strongly. Similar objections can be raised
about other cases, except for cases 2,7,8. Case 2 with $\epsilon=0.21,
\delta=.14$ yields
\begin{equation}
\left[\begin{array}{ccc}
2.10^{-6} & 0.001 & 6.10^{-3} \\
0.001 & 0.03 & 6.10^{-3} \\
6.10^{-5} & 0.03 & 1\\ \end{array}\right],\label{1mod2}
\end{equation}
which fits Eq.\ref{mod1} well apart from a factor $\sim$3 in the 12
position. 
The next sub-dominant operator in the 22 position needs to be
$2.10^{-3}$ according to Table~\ref{tab:res1}. The values of
$\epsilon$ and $\delta$ used in Eq.~\ref{1mod2} give the subdominant
operator in the 22 position to be $\sim 6.10^{-3}$. This is
acceptable, but a closer match occurs for the next higher dimension
operator, which has magnitude $\sim 10^{-3}$. An ambiguity occurs in
that we have
not set the normalisation of the sub-dominant operator due to its
numerous possibilities and so the original discrepancy factor of $\sim
3$ could easily be explained. Below, we do not consider the numerical
size of the
sub-dominant operator because it is clear that some operator can be
chosen that will fit the required number well.
If the charge assignments under the U(1)$_X$ symmetry were the same as
in this case, we would have succeeded in explaining why the assumption
of texture zeros was valid. For example, the 13
element in Eq.\ref{1mod2} being $6 \times 10^{-3}$ instead of zero only
affects mixing angle and mass predictions by a small amount. We have
also explained the hierarchies between the elements in terms of the
different mass scales involved in the non-renormalisable operators by
not having to choose dimensionless parameters of less than 1/3 (or
greater than 3).
Case 7 with $\epsilon=0.36, \delta=0.08$ gives
\begin{equation}
\left[\begin{array}{ccc}
2.10^{-6} & 0.001 & 4.10^{-3} \\
0.001 & 0.01 & 4.10^{-3} \\
6.10^{-5} & 0.03 & 1\\ \end{array}\right]. \label{1mod7}
\end{equation}
We should note that at this level, we may naively expect 8\%
corrections to the constraint in Eq.\ref{1mod7} through the next order
of $\delta$
operators in each element. We could have attempted to include these
possible errors in the numerical fits but we did not due to the fact
that they are very model dependent. Deeper model building in terms of
constructing the non-renormalisable operators out of extra fields or
examining underlying string models would be required to explain why
this should not be the case. It should also be borne in mind that
explanations for exact texture zeroes can be made in this context by
setting fractional $U(1)_X$ - charges on 
the heavy fields in the operators, or by
leaving certain heavy fields out of the FN model. Case 8 with
$\epsilon=0.36, \delta=0.08$ gives the same results as in
Eq.\ref{1mod7}, except with the (22) element as 0.03.

{}From Table~\ref{tab:res1} we see that model 3 (the model that fits the
data the best) has
\begin{equation}
H_{ij}^{emp} \sim \left[\begin{array}{ccc}
0 & 0.002 & 0 \\
0.002 & 0.03 & 0\\
0 & 0.03 & 1\\ \end{array}\right]. \label{mod3}
\end{equation}
Choosing $\epsilon=0.26, \delta=0.12$ in case 2 gives a good match to
Eq.\ref{mod3}:
\begin{equation}
\left[\begin{array}{ccc}
9.10^{-8} & 0.002 & 8. 10^{-3} \\
0.002 & 0.03 & 8.10^{-3} \\
6. 10^{-5} & 0.03 & 1\\ \end{array}\right]. \label{3mod2}
\end{equation}
Case 7 with $\epsilon=0.40, \delta=0.07$ or case 8 with the same
$\epsilon$ and $\delta$ both give a fairly good match as well.

Model 4 is different in the sense that it possesses a hierarchy
between the 12 and 21 entries of the effective Yukawa couplings:
\begin{equation}
H_{ij}^{emp} \sim \left[\begin{array}{ccc}
0 & 0.0007 & 0 \\
0.004 & 0.03 & 0\\
0 & 0.03 & 1\\ \end{array}\right]. \label{mod4}
\end{equation}
Here, case 1 with $\delta=0.2, \epsilon=.15$ predicts
\begin{equation}
\left[\begin{array}{ccc}
3.10^{-7} & 0.0007 & 4. 10^{-3} \\
0.004 & 0.03 & 4.10^{-3} \\
 10^{-5} & 0.03 & 1\\ \end{array}\right], \label{3mod1}
\end{equation}
an extremely good match to Eq.\ref{mod4}. Case 6 with $\epsilon=0.28,
\delta=0.11$ provides a good match also. 

Thus we see that we can explain the hierarchies and texture zero
structures of the models that fit the data best. In general, it seems
likely that we have enough freedom in setting charges to attain the
required hierarchies for the Yukawa matrices.

\section{The String Model}
 
In the following, we will present a semi-realistic string model
which provides an existence proof of how previously described
non-renormalisable operators 
may be generated from first principles using string theory. 
Before this, let us briefly comment on how the basic features of the
$U(1)_X$ symmetries that we have discussed arise in string constructions.

In realistic free fermionic 
string models \cite{fc,leo2}  there are 
some general features: 
At a scale $M_{string}\sim  5 g_{string}\times 10^{17}$GeV, one obtains
an effective $N=1$ supergravity model with a gauge symmetry structure which
is usually a product of non-Abelian groups times several $U(1)$ factors.
The non-Abelian symmetry contains an observable and a hidden sector.
The massless superfields accommodating the  higgs and known chiral
fields transform non-trivially under the observable part and usually
carry non-zero charges under the surplus $U(1)$-factors. The latter,
act as family symmetries in the way described above.
Some  of them are anomalous, but it turns out that one can usually
define new linear $U(1)$ combinations where all but one are anomaly-free.
The anomalous $U(1)$ is broken by the 
Dine Seiberg Witten mechanism \cite{dsw},
in which a potentially large Fayet-Iliopoulos D-term  is generated  by
the VEV of the dilaton field. A D-term however  breaks supersymmetry 
and destabilizes the string vacuum, unless there is a direction in  the scalar
 potential which is D-flat and F-flat with respect to the  non-anomalous 
gauge symmetries. If such a direction exists, some of the singlet fields
will acquire a VEV,  canceling the anomalous D term, so that  supersymmetry 
is restored.
Since the fields corresponding to such a flat direction typically also
carry charges for the non-anomalous D-terms, they  break all $U(1)$
symmetries spontaneously.
For the string model in ref.\cite{leo2},
the expected order of magnitude for
the VEV of the singlet fields is $<\Phi_i>\sim (0.1-0.3)\times M_{string}$.
Thus, their magnitude is of the right order to produce the required  mass 
entries in the mass matrices via non-renormalisable  operators.

As an application of the above procedure, we will make a
first  attempt to derive the relevant operators
for the mass matrices of the  model
based on the work in ref.\cite{leo2}.
The  string model is defined in terms of nine basis vectors
 $\{S,b_1,b_2,b_3,b_4,b_5,b_6,\alpha,\zeta\}$ and a suitable
choice of the GSO projection coefficient matrix.
The resulting gauge group has a Pati-Salam ($ SU(4)\times
SU(2)_L\times SU(2)_R$) non-Abelian  observable part,
accompanied by four $U(1)$  Abelian factors and
a hidden $SU(8)\times U(1)$ symmetry.
 
In the following, for convenience, we
denote a set of complex right fermions with the letters
$\{\bar\Psi^{1\cdots 5}, \bar\varphi^{1\cdots 6},\bar\eta^{123},
\bar{z}^{12}\}$  and real right fermions with
 $\{\bar y^{1\cdots 6},\bar\omega^{1\cdots 6}\}$.
Now, a specific model is defined in terms of a set of
boundary conditions on the phases picked up by the
fermions when parallel transported around non-contractible
loops.  The model
is derived from the following basis\cite{leo2}:
\begin{eqnarray}
S & = &\{\psi^{\mu},\chi^{12\cdots 6} \; \; \; ;
0\cdots 0\}
\nonumber\\
b_1 & = & \{\psi^{\mu},\chi^{12},y^{3456}\bar{y}^{3456} \; \; \; ;
\bar\Psi^{1\cdots 5}\bar\eta^1\}
\nonumber\\
b_2& = & \{\psi^{\mu},\chi^{34},y^{12}\bar{y^{12}}
\omega^{56}\bar\omega^{56} \; \; \; ;
\bar\Psi^{1\cdots 5}\bar\eta^2\}
\nonumber\\
b_3& = &\{\psi^{\mu},\chi^{56},\omega^{1234}\bar\omega^{1234} \; \; \; ;
\bar\Psi^{1\cdots 5}\bar\eta^3\}
\nonumber\\
b_4& = &\{\psi^{\mu},\chi^{12},y^{36}\bar{y}^{36},
\omega^{45}\bar\omega^{45} \; \; \; ;
\bar\Psi^{1\cdots 5}\bar\eta^1\}
\nonumber\\
b_5& = &\{\psi^{\mu},\chi^{34},y^{26}\bar{y}^{26},
\omega^{15}\bar\omega^{15} \; \; \; ;
\bar\Psi^{1\cdots 5}\bar\eta^2\}
\nonumber\\
b_6& = &\{0,0,y^{6}\bar{y}^{6},
\omega^{15}\bar\omega^{15} \; \; \; ;
\bar\Psi^{1\cdots 5}\bar\eta^{123}\bar\varphi^{123}\bar{z}^1\}
\nonumber\\
\zeta & = &\{0,\cdots 0 \; \; \; ; \bar{z}^{12}\bar\varphi^{1\cdots 6}\}
\nonumber\\
\alpha & = & \{y^{46}\bar{y}^{46},
\omega^{46}\bar\omega^{46} \; \; \; ;
\bar\Psi^{123}\bar\eta^{12}\bar{z}^{12}\}
\nonumber
\end{eqnarray}
All world sheet fermions appearing in the basis are assumed to
have periodic boundary conditions, while those not appearing
are anti-periodic. An immediate consequence of using only {\it periodic}
\/and {\it anti-periodic} \/boundary conditions is that the resulting gauge
symmetry is in general a product of $SO(n)$ groups. Thus, in the 
above basis for example, the complex world sheet fermions
$\bar\Psi^{1,\ldots ,5}$
define an $SO(10)$ symmetry which is broken by the last vector $\alpha$
into $SO(6)\otimes O(4)$. Now, bearing in mind that this part will be
interpreted as the observable gauge symmetry, we observe the isomorphies
$SO(6)\sim SU(4)$, $O(4)\sim SU(2)\otimes SU(2)$. The two $SU(2)$s are 
going to accommodate the left and right components of the
matter fields. Thus, the resulting gauge symmetry is isomorphic 
to the Pati-Salam gauge group. Thus the complete symmetry of the
model under the 
above choice is
\begin{eqnarray}
[SU(4)\times SU(2)\times SU(2)\times U(1)^3]_{o}\times
[SU(8)\times U(1)]_{h}
\end{eqnarray}
where the subscripts ${(o,h)}$ denote the observable and the hidden
part respectively. With the specific choice of the projection
coefficient matrix in \cite{leo2}, one obtains three chiral
families in the $(4,2,1)+(\bar 4,1,2)$ representations
of the PS symmetry,
and two higgs pairs transforming as $(4,1,2)+(\bar 4,1,2)$,
all arising from the sectors $b_{1,2,3}$
and $b_4,b_5$.

In particular, the  massless spectrum contains
three $(F_{1,3,4})_L = (4,2,1)$ representations obtained from the
sectors $b_{1,3,4}$, which accommodate the left handed fermion fields.  
There are five $(\bar 4,1,2)$ representations
$(\bar{F}_{1,4,5},\bar{F}_2,\bar{F}_2')_R$
named after the corresponding  sectors  and
two $H_{4,5}=(4,1,2)$ arising from the sectors $b_{4,5}$.
Thus, two linear combinations of the $\bar{F}_i$ will
play the role of the GUT higgs $\bar{H}$, while the
remaining three $\bar{F}$'s accommodate the
right- handed fermions. The spectrum includes also bidoublets
$h_i=(1,2,2)_i$, sextets $D_i=(6,1,1)_i$ and a sufficient number
of singlet fields $\Phi_{ij},\xi_j,\zeta_k$.
A certain number of  singlets should develop VEVs
in order to satisfy the flatness conditions and
give masses to unwanted colour triplets and exotic
states.

 In addition, one obtains fractionally charged
states which arise in non-standard representations
of the PS- symmetry, namely $(1,1,2)$, $(1,2,1)$ and
one pair $(4,1,1)+(\bar 4,1,1)$. {}Finally, under the
hidden gauge group, one obtains 10 irreducible
representations $Z_i,\bar{Z_i}$ sitting
in the {\underline 8} of SU(8) while carrying quantum
numbers under all five $U(1)$ symmetries of the model.
All  states are divided to
those arising from the Neveu-Schwarz (NS) and Ramond (R)
 sectors. In particular
 the NS-sector gives the graviton multiplet as well
as the singlet fields $\Phi_i,\Phi_{ij}$, sextets and
the bidoublets $h_{3},\bar h_3$.

\section{Calculation of Tree Level and Non Renormalisable Operators
in the  String Model}
 
To calculate the superpotential of the model, one needs to
obtain vertex operators for all physical states of the theory.
To construct vertex operators for the states of a given model, every
world- sheet fermion has to be represented by a conformal field.
In the case that a representation of the model can be fully
factorized in a left and a right moving piece, one can pair
them up to bosonized fields.
Now, according to the definition of the supersymmetry generator $S$
in the above basis of our model, one can conclude that the left moving
fields $\chi^i$ can be bosonized $(\chi^1\pm \imath\chi^2)/\sqrt{2}
= \exp\{\pm \imath S_{12}\}$ and similarly for the $\chi^{3,4}$
and $\chi^{5,6}$ pairs.
$N=1$ supersymmetry implies the existence of an extra  current,
 which is expressed in terms of $S_{ij}$ as  follows \cite{kln}
\beq
J(q)= \imath \partial_q (S_{12}+S_{34}+S_{56})
\eeq
and which is extended to three $U(1)$'s generated
by $S_{12},S_{34},S_{56}$.
 
The Yukawa couplings in four dimensional 
superstring models correspond to
expectation values of the form
\beq
<\int d^2q_1\int d^2q_2 \int d^2q_3 V_1^F(q_1)V_2^F(q_2)V_3^B(q_3)>
\eeq
where the $V_i^{F,B}$ are the vertex operators for the fermionic (F) and
the bosonic (B) states, while $q_{1,2,3}$ are the two dimensional
coordinates.
Thus, a vertex operator for any physical state is a collection of
conformal fields that represent the quantum numbers of the state under
 all symmetries of the model.
The piece of the vertex operators involving the bosonized left moving
fields $\chi^i$ is given for the bosons
 by $V^B_{-1}\sim \exp\{\alpha S_{12}\}
\exp\{\beta S_{34}\}\exp\{\gamma S_{56}\}$. Similarly, for the fermions,
$V^F_{-1/2}\sim \exp\{(\alpha -1/2) S_{12}\}
\exp\{(\beta -1/2) S_{34}\}\exp\{(\gamma -1/2) S_{56}\}$.
The subscripts $-1,-1/2$ refer to the
corresponding ghost numbers.
The total ghost number should add up to $-2$, thus in trilinear terms
the non-vanishing couplings are proportional to the correlator
$<V^FV^FV^B>$. In non-renormalizable contributions, the remaining
vertex operators $V^B_4\cdots V^B_n$ have to be `picture - changed'
in the zero picture \cite{kln}. In general,  a particular correlator
is  non - vanishing, only if it is invariant under the three $U(1)$'s.
In addition it has to respect the usual (right moving) gauge invariance
and other global symmetries. {}For example, pure NS- couplings are possible
only at tree level. The same is true for higher order couplings
involving only Ramond fields, and so on. A complete list
of rules is found in \cite{kln}.
 
If we imply the well defined set of rules to calculate
the Yukawa interactions in
the present string model, we obtain the following
tree-level terms that are relevant to our discussion
 
 
\begin{eqnarray}
{\cal{W}}& \rightarrow & F_{4L}\bar{F}_{5R}h_{12}+\frac 1{\sqrt{2}}
             F_{4R}\bar{F}_{5R}\bar\zeta_2+\bar{F}_{3R}F_{3L}h_3
\nonumber\\
&+& \bar\xi_1h_3h_{12}+\xi_4h_3\bar{h}_{12}+
 \bar\Phi_{12}h_{12}h_{12} + \Phi_3\bar{h}_{12}h_{12}
\nonumber\\
&+&\xi_1\bar h_3\bar h_{12}+\bar\xi_4\bar{h}_3{h}_{12}+
\Phi_{12}\bar{h}_{12}\bar{h}_{12} +
\bar\Phi_3{h}_{12}\bar{h}_{12}
+\cdots \label{sup}
\end{eqnarray}
where the $\{\cdots\}$ stand for terms involving exotic and hidden
fields and other couplings irrelevant for our purpose. The
F-flatness conditions are derived for the complete tree- level
superpotential,
which is given in  \cite{leo2} and involves in total 18
singlet fields. Five of these fields, namely $\Phi_{1,\ldots,5}$ 
have zero quantum numbers under the $U(1)$ groups,
while the rest of the fields (denoted by $\xi_{1,2,3,4}$,
$\bar{\xi}_{1,2,3,4}$, $\zeta_{1,2}$,
$\bar{\zeta}_{1,2}$, $\Phi_{12}$,
$\bar{\Phi}_{12}$,  $\Phi_{\overline{12}}$,  $\bar{\Phi}_{
\overline{12}}$,) have non-trivial quantum numbers.
 
{}From the above, it is clear that only a few Yukawa couplings
are available for fermion mass generation at the tree level.
The missing terms are expected to be obtained from
non-renormalisable (NR)  terms. In the case of the PS
- symmetry we expect NR-terms  of the form
\begin{equation}
\bar{F}Fh\frac{\bar{H}H\Phi_i\Phi_j}{M^4_{string}},{\cdots}\;\; etc
\label{snr}
\end{equation}
which act as effective mass operators once the fields
$H$, $\bar{H}$, and $\Phi_{i,j}$ get VEVs. The scale
where the higgs  fields $H,\bar H$  obtain their VEVs
is determined from phenomenological requirements
and renormalisation group analysis\cite{rge} of the particular model.
Moreover, the singlet VEVs are
not completely  arbitrary since they should satisfy
the $D$- and $F$- flatness conditions. In general,  the
$D$-flatness conditions read
\begin{eqnarray}
\sum_iQ_X^i\mid <\Phi_i>\mid^2+
\frac{g^2}{192\pi^2}\mbox{Trace}\{Q_{U(1)_X}\}M_{Pl}^2=0\label{u1an}\\
\sum_iQ_n^i\mid <\Phi_i>\mid^2=0 \label{u1nan}
\end{eqnarray}
where $<\Phi_i>$ are the singlet VEVs and $g$ stands for the
unified gauge coupling at $M_{string}$. $U(1)_X$ in
 (\ref{u1an}) is the anomalous $U(1)$ combination and $Q_X^i$
the corresponding $U(1)_X$ charge of the singlet $\Phi_i$.  Eq.
(\ref{u1nan}) holds for all the non- anomalous $U(1)$ symmetries
of the particular model.  From the relations (\ref{u1an},\ref{u1nan}),
it is clear  that the order of magnitude of
 the VEVs of the singlet fields is determined by the Trace - term.
Thus, we expect that
\begin{equation}
<\Phi>^2\sim
{\cal O}\left(\frac{g^2\mbox{Tr}(Q_X)}{192\pi^2}\right)M_{Pl}^2
\end{equation}
In particular, for the string model in ref. \cite{leo2},
Tr$[Q_X]=72$, therefore the order of magnitude for
the singlet fields is $<\Phi_i>\sim (0.1-0.3)\times M_{string}$.
(See also appendix 3 for the details.)
This indicates that the singlet VEVs
have the correct magnitude, in order to produce the required mass
entries in the mass matrices  via the non - renormalisable  operators
of Eq.(\ref{snr}).
We also note here that the spontaneous breaking
of the anomalous U(1) symmetry introduces one more mass scale
$M_X$ in the theory, which is characterised by the magnitude of
the related singlet VEVs. Thus, one naturally expects the hierarchy
$M_{string}\ge M_{X}\ge M_{GUT}$.

One possible choice of non-zero VEVs is
\begin{equation}
<\bar\Phi^-_{12}>,<\Phi_{12}>,<\xi_1>,<\bar\xi_2>
\label{vevs}
\end{equation}
and $<Z_5>,<Z'_8>\not= 0$ of the hidden fields. Solving
the flatness conditions (appendix 3), one finds that the order
of magnitude of the singlet VEVs is $\sqrt{\frac{a_u}{\pi}}$
in Planck units. It is easy to see that the choice (\ref{vevs})
satisfies trivially the $F$ flatness conditions.
We should point out however that this choice is not unique.
There are other cases which also satisfy conditions
(\ref{u1an},\ref{u1nan}) and hopefully, a
solution which meets the phenomenological requirements
does exist.
The non- zero VEVs in (\ref{vevs}) provide
all dangerous colour triplets with masses from tree-level superpotential
terms. Here we would like to investigate if they are
also capable of producing the relevant operators
for the fermion masses. This computation will prove to be a rather
hard task mainly due to the rapidly increasing number
of NR-operators as the calculation proceeds to higher
orders. We will see however, that the pattern of
the fermion mass matrices described in the previous
sections is basically obtained.

We will first start the examination of the tree level superpotential.
Due to the string symmetries and the $U(1)$ charges
of the superfields, as can be seen from (\ref{sup})
only three terms relevant to the
fermion masses exist at three level
\begin{equation}
{\cal{W}} \rightarrow F_{4L}\bar{F}_{5R}h_{12}+\frac 1{\sqrt{2}}
             F_{4R}\bar{F}_{5R}\bar\zeta_2+\bar{F}_{3R}F_{3L}h_3
\label{tls}
\end{equation}
 Here, $h_{12},h_3$ are bidoublets, $\bar\zeta_2$ is a singlet,
while the $F_{L,R}$ chiral fields have been presented previously.
We may give a non zero VEV to one of the two bidoublet
 higgs fields
(or to  a linear combination $\cos\theta h_{12}+\sin\theta h_3$)
and support one generation with masses at the tree level.
Since there are more than one doublets in the spectrum,
first, we should determine the massless state along
the chosen flat direction. At the tree level, the bidoublet
higgs mass matrix obtained from the relevant terms is
\begin{equation} (h_{3} , h_{12},
                      \bar{h}_{3}, \bar{h}_{12} )
\left(\begin{array}{cccc}
   0 &\bar\xi_1&0&\xi_4 \\
   \bar\xi_1&\bar\Phi_{12}&\bar\xi_4 & \varphi_3\\
   0&\bar\xi_4&0&\xi_1\\
   \xi_4&\varphi_3&\xi_1&\Phi_{12}
   \end{array} \right)
\left(\begin{array}{c} h_{3} \\ h_{12}\\
                      \bar{h}_{3}\\ \bar{h}_{12}\end{array}\right)
\end{equation}
with $\varphi=\Phi_3/2$.
In order to have at least one non-zero  eigenvalue we
impose $\det[m_h]\equiv (\xi_1\bar{\xi}_4-\bar{\xi}_1\xi_4)^2=0$,
which is satisfied for any value of the $\Phi_3,\Phi_{12},
\bar\Phi_{12}^-$ VEVs, provided $\xi_1\bar{\xi}_4=\bar{\xi}_1\xi_4$.
The choice (\ref{vevs}) is consistent with these requirements.
Moreover, it leaves $h_3,h_{12}$ massless at three level.
We then let $h_{12}$ develop a VEV and give masses to
the top, bottom and tau particles living in the $F_{4L},
\bar{F}_{5R}$ representations. The $ h_3$ bidoublet is expected
to receive a mass from a NR- term. Thus, to proceed
further, we need the contributions of the non-renormalisable
terms. As in the tree level case, a non-vanishing NR- term of
 the superpotential must obey all  the string selection rules
 \cite{kln} and be invariant under all the gauge and global
symmetries. Since here we discuss the fermion masses,
we are primarily interested in those operators contributing
to the corresponding matrices. At fourth order, we find
no relevant terms.
At fifth order, there are several operators which in
principle could contribute to the fermion mass matrices.
We list them here,
\begin{eqnarray}
\bar{F}_{3R}F_{3L}h_3\xi_1\bar\xi_2\;\; ,&
\bar{F}_{3R}F_{3L}h_3\Phi_{3,4,5}^2 
\label{5.}\\
\bar{F}_{1R}F_{1L}\bar{h}_{12}\bar\zeta_2\Phi_2\;\; , &
\bar{F}_{5R}F_{4L}h_{12}\Phi^2_{1,2}
 \label{5a} \\
F_{5R}F_{4L}\bar{F}_{2R}\bar{F}_{2R}'h_{12}&
\label{5b}
\end{eqnarray}
(scaled with the proper powers of $M_{string}$). Let us
analyse the above contributions in terms of the 
particular flat direction chosen here.
It is clear that, irrespective of the choice of the
singlet VEVs, the terms (\ref{5.}) do not add a new
contribution since they constitute small corrections to
the already existing tree-level term $\bar{F}_3F_3h_3$.
Moreover, within the given choice of our flat direction,
$<\Phi_i> =<\bar\zeta_2> = 0$, the terms (\ref{5a}) do not
also generate any new fermion mass term.  Thus, there is only
one term which contributes to the fermion mass matrices,
namely the operator of Eq.(\ref{5b}).
This is a  $n=1$ operator according
 to our  classification in the  previous sections.
 We have already interpreted $\bar F_{5R}$ and
$F_{4L}$ as the left and right components of the
third fermion generation.
Up to now we have not determined which of $\bar{F}_{2R}$,
$\bar{F}'_{2R}$ is going to play the role of the second family.
The $5^{th}$ order operator still leaves this undetermined
since both fields enter in the
operator in a symmetric way.
Thus there are two options.
Either we set $<\bar{F}_{2R}> = 0$, or we have to rename the fields
so that $F_{3L},\bar{F}_{3R}$ are the 3$^{rd}$ generation
fermions and $h_3$  the massless higgs.
In the first case we retain the 5th order
contribution to the mass matrix, while in the second we have
a unique choice for the 2$^{nd}$ family and the higgs, i.e.,
$\bar{H} = \bar{F}'_{2R}$ and $\bar{F}_{2R}$
accommodates the right-handed fields of the second generation.
 
In order to calculate the contribution
of the operator in eq.(\ref{5b})
to the mass matrices,
we should properly contract the various fields
involved  in the NR-term. In principle, the
numerical coefficient in front of the desired
operator is a linear combination of the
Clebsch Gordan coefficients presented in the Table,
each of them multiplied by a different phase
factor. Our ignorance about the numerical
coefficients of the mass matrix entries has been
minimized in the unknown phase factors.
However,  we can make a natural assumption
that the biggest contributions come from contractions
occurring
first for the fields belonging to the same sector.
Recalling now  that $\bar{F}_{2R}$ and $\bar{H}
\equiv \bar{F}'_{2R}$
originate from the same sector $b_2$, while $F_{4L}$,
$H\equiv F_{5R}$ and $h_{12}$ are obtained from $b_{4,5}$.
We find that
\begin{eqnarray}
(F_{4L}H) (\bar{F}_{2R}\bar{H})h_{12}&\ra & O^G\nonumber\\
       &  \ra &\{\frac{2}{\sqrt{5}}Qd^ch_d,\frac{4}{\sqrt{5}}\ell
          e^ch_d\} \label{d23}
\end{eqnarray}
 i.e., this operator contributes to
 down quark and charged lepton mass
matrices. Since this contribution is the
second largest after the tree level term $F_{4L}\bar{F}_{5R}h_{12}$,
 we identify (\ref{d23}) with the 23 - entry of
the corresponding mass matrices. It is clear therefore,
that in this picture $F_{5R},\bar{F}_{2R}$ accommodate
the right components of the 3$^{rd}$ and 2$^{nd}$ generations
respectively, whereas $F_{4L}$ contains the left fermions
of the heavy generation.
In order to obtain non-zero
s-quark and $\mu$- masses, we need to fill in the
32 entry of the down and charged lepton mass matrices
with a higher order operator, so that the $2\times 2$
lower block of the corresponding mass matrices exhibits
a structure of the asymmetric type considered in the previous section,
\beq
{\cal F_R}{\cal M}_{d,e}{\cal F_L}=(\bar{F}_{2R},\bar{F}_{5R})
\left(\begin{array}{cc}
   \eta_{22} & <H\bar{H}> \\
      {\eta_{32}}  & 1\end{array} \right) h_{12}
\left(\begin{array}{c} F_{iL} \\ F_{4L}\end{array}\right)
\label{stringblock}
\eeq
where $\eta_{ij}$ stand for  higher order NR- contributions,
while $F_{iL}$ represents in general one of the two remaining
left-handed fourplets $F_{(1,3)L}$. To determine which of
the latter will accommodate the 2$^{nd}$ generation and
calculate $\eta$'s, one has to proceed above the fifth order and
find the relevant non-vanishing correlators. For example, choosing
a new flat direction  in which $\bar{\zeta}_2,\Phi_2$ singlets develop
non-zero VEVs while interpreting $F_{3L},\bar{F}_{3R}$ as the third
generation, $\eta_{22}$ may arise from the fifth order NR-operator
$F_{1L}\bar{F}_{1R}\bar{h}_{12}\bar\zeta_2\Phi_2$ in Eq.(\ref{5.}).
Interestingly, this is a $n=0$ operator according
 to our classification, however it is suppressed  compared to
 a tree level term due to the presence of the `effective' flavour 
 factor  $\delta^2=\frac{<\bar\zeta_2\Phi_2>}{M_{string}^2}$.  Furthermore, 
 higher NR-terms will certainly involve $n=2,3$ etc operators. 
 Thus,  it is  clear that the above procedure will require $n\ge  2$
operators and the analysis will be more involved than  the field theory
model described in the our earlier sections.

As a matter of fact, a detailed analysis requires also the examination of all
possible flat directions as well as calculation of the non-renormalizable
contributions to even higher orders, since one has to ensure that
the necessary Weinberg-Salam doublets living in some combination
of our $h_3,h_{12}$ bidoublets remain massless up to this order.
{}For the moment, in our first approach to this model,
we have been able to show that the rather complicated string
construction  stays in close analogy with the field theory approach
presented in the early sections.


\section{Conclusions \label{conclusion}}

We have examined Yukawa textures within a string inspired  $SU(4)\times O(4)$ 
model extended by a gauged $U(1)_X$ family symmetry and non-renormalisable
operators above the unification scale of the form  in Eq.\ref{newoperators}.
These operators factorise into a  factor $(H\bar{H})$ and a factor involving 
the singlet fields $\theta,\bar\theta$. The singlet fields $\theta,\bar\theta$ 
break the $U(1)_X$ symmetry and provide the horizontal family hierarchies
while the $H,\bar H$ fields break the SU(4)$\otimes$SU(2)$_L\otimes$SU(2)$_R$
symmetry and give the vertical  splittings arising from group theoretic
Clebsch relations between different charge sectors.  The factor $(H\bar{H})$
also provides an additional flavour independent suppression factor $\delta$
which  helps the fit. The quark and lepton masses and quark mixing angles
are thus described at high energies by a single unified Yukawa matrix
whose flavour structure is controlled by a broken $U(1)_X$ family
symmetry, and all vertical splittings controlled Clebsch factors.
An important feature of the scheme is the existence of Clebsch zeroes
which allow an entirely new class of textures to be obtained.
For example the RRR solutions 3 and 5 may be reproduced by this scheme
which are complementary to the RRR solution 2 favoured
by the IR approach.

In addition to the symmetric textures we have also
performed a completely new analysis of the non-symmetric textures
which are motivated by the string construction.
A global fit to the fermion mass spectrum 
with 3 dof is described, in which
three models in Table~\ref{tab:res1}
are singled out with $\chi^2$/dof$<1$.\footnote{By comparison a recent
paper~\cite{stats} performed a global $\chi^2$
analysis for some $SO(10)$ models, including the mass and mixing data.
With 3 dof, they obtain a $\chi^2$/dof$\sim1/3$ for the best model.
While our fit to model 3 in Table~\ref{tab:res1}, for example,
has a smaller $\chi^2$/dof than this, it 
is difficult to make a comparison as
in ref.\cite{stats} quark mass correlations from data, as
well as the effect of large $\tan\beta$ on $m_{b}$ has been
included. Also note that these involve the soft terms, thus a larger number of 
parameters are involved in the fit.}
At this level of difference of $\chi^2$ between models, the
$\chi^2$ test is subject to large statistical fluctuations.
Therefore, we do not statistically distinguish between the fits in
Tables~\ref{tab:res1},\ref{tab:res2} since both contain good fits to
the data with $\chi^2$/dof$ < 1$. However we have a theoretical preference
for the models in Table~\ref{tab:res1} since these models
result from the operators
in Eq.\ref{newoperators} where the family hierarchies are accounted for
by the $U(1)_X$ symmetry, as explained in section~\ref{familysymmetry}. 
By contrast, the models in Table~\ref{tab:res2} 
result from the operators in Eq.\ref{op} and are 
essentially an up-dated version of those previously considered in 
ref.\cite{422}.

The string analysis performed in the later sections of the paper
lends some support to the approach followed in this model.
In the  string model,
the $U(1)$ family symmetries are a consequence of the string
construction, but there are four of them with one being anomalous.
There are several singlets (charged under the family group)
to take the role of the $\theta$ fields and the 
$n=1$ operators involving a factor of $H\bar{H}$
are clearly expected in the effective theory below the string scale.
We have shown that operators such as
$O^G$ which were simply pulled out of thin air in the earlier parts 
of the paper may in fact originate
from string theory. 
As an example we constructed explicitly the lower $2\times 2$ block
in Eq.\ref{stringblock} which has the characteristic asymmetric structure
of the Yukawa textures considered earlier.
It will be noted however that the lower $2\times 2$ block 
in Eq.\ref{stringblock} does not
correspond precisely to the ansatz in Eq.\ref{A_1}. 
Within the given string construction, such an ansatz
does not appear to be possible.
The reason is the extra $U(1)$ symmetries and the other
discrete-like symmetries (selection rules) left over
in the low energy model. 
A new string construction with a new boundary condition
on the string basis is required in order to make contact with
the phenomenologically preferred ansatze.
This will be the subject of future work.

\begin{center}
{\Large {\bf Acknowledgments}}
\end{center}
B.A. would like to thank J.Holt for advice on the $\chi^2$ test.
The work of S.F.K. is partially supported by PPARC grant number GR/K55738.
S.L. would like to acknowledge the Theory Group at the University of 
Southampton for an one-month PPARC funded Research Associateship
which greatly facilitated this research.
The work of S.L. is funded by a Marie Curie Fellowship
(TMR ERBFMBICT-950565) and at the initial stages by
$\Pi$ENE$\Delta$-91$\Delta$300. 
The work of G.K.L. is
partially supported by $\Pi$ENE$\Delta$-91$\Delta$300. 

\newpage

\begin{center}
{\bf Appendix 1. $n=1$ Operators}
\end{center}

The $n=1$ operators are by definition all of those operators 
which can be
constructed from the five fields $F \bar{F} h H \bar{H}$ by
contracting the group indices in all possible ways, as discussed in
Appendix 1. After the Higgs fields $H$ and $\bar{H}$
develop VEVs at
$M_{GUT}$ of the form $\langle H^{\alpha b} \rangle = \langle H^{41}
\rangle = \nu_H$, $\langle \bar{H}_{\alpha x} \rangle = \langle 
\bar{H}_{41} \rangle = \bar{\nu}_H$, the operators listed in the
appendix yield effective low energy Yukawa couplings with small 
coefficients of order $M_{GUT}^2/M^2$. However, as in the simple example
discussed previously, there will be precise Clebsch relations between
the coefficients of the various quark and lepton component fields.   
These Clebsch relations are summarised in Table~\ref{tab:clebschn1},
where  relative normalisation factor has been applied to each.
The table identifies which SU(4) and SU(2) structures have been used
to construct each individual operator by reference to
Eqs.~\ref{su4str},\ref{su2str}.

\begin{table} 
\begin{center}
\begin{tabular}{|c|c|c|c|c|c|c|} \hline
 & SU(2) & SU(4) & $Q \bar{U} h_2$ & $Q \bar{D} h_1$ & $L \bar{E} h_1$
& $L \bar{N} 
h_2$
\\ \hline
$O^A$ &I & I &1 & 1 & 1 &1 \\ 
$O^B$ &II & I& 1 & -1& -1 &1 \\ 
$O^C$ &I & II &  $\frac{1}{\sqrt{5}}$ &
$\frac{1}{\sqrt{5}}$ &$\frac{-3}{\sqrt{5}}$ & $\frac{-3}{\sqrt{5}}$  \\ 
$O^D$ &II & II& $\frac{1}{\sqrt{5}}$ & $\frac{-1}{\sqrt{5}}$
& $\frac{3}{\sqrt{5}}$ & $\frac{-3}{\sqrt{5}}$\\ 
$O^E$ &III & III &  0 & 2 & 0 & 0 \\
 $O^F$ &II & III & $\sqrt{2}$ & $-\sqrt{2}$& 0 & 0\\
$O^G$ &III & IV & 0 & $\frac{2}{\sqrt{5}}$ & $\frac{4}{\sqrt{5}}$ & 0 \\
 $O^H$ & IV & IV & 4/5 & 2/5 & 4/5 & 8/5 \\
 $O^I$ & V & V& 0 & 0 & 0 & 2\\
$O^J$ & VI & V & 0 & 0 & $\frac{4}{\sqrt{5}}$ & $\frac{2}{\sqrt{5}}$\\
$O^K$ & V & VI & 8/5 & 0 & 0 & 6/5 \\
$O^L$ & IV & VI & $\frac{16}{5 \sqrt{5}}$ & $\frac{8}{5 \sqrt{5}}$ &
$\frac{6}{5 \sqrt{5}}$ & $\frac{12}{5 \sqrt{5}}$ \\
$O^M$ & III & I &0 & $\sqrt{2}$ & $\sqrt{2}$ &0 \\ 
$O^N$ &V & III&2&0&0&0\\ 
$O^O$ & V& IV&$\frac{2}{\sqrt{5}}$ & 0 & 0 & $\frac{4}{\sqrt{5}}$ \\
$O^P$ & I & VI & $\frac{4 \sqrt{2}}{5}$ & $\frac{4 \sqrt{2}}{5}$ & 
$\frac{3 \sqrt{2}}{5}$  & $\frac{3 \sqrt{2}}{5}$  \\
$O^Q$ & II & VI & $\frac{4 \sqrt{2}}{5}$ & -$\frac{4 \sqrt{2}}{5}$ & 
-$\frac{3 \sqrt{2}}{5}$  & $\frac{3 \sqrt{2}}{5}$ \\
$O^R$ & III&VI&0&$\frac{8}{5}$& $\frac{6}{5}$& 0 \\
$O^S$ & VI & VI & $\frac{8}{5 \sqrt{5}}$ & $\frac{16}{5 \sqrt{5}}$ &
$\frac{12}{5 \sqrt{5}}$ & $\frac{6}{5 \sqrt{5}}$ \\
$O^T$ &IV & I&  $ \frac{2 \sqrt{2}}{5}$&  $ \frac{ \sqrt{2}}{5}$&  $ \frac{
\sqrt{2}}{5}$&  $ \frac{2 \sqrt{2}}{5}$\\
$O^U$ &VI&I& $ \frac{ \sqrt{2}}{5}$&  $ \frac{2 \sqrt{2}}{5}$&  $ \frac{2
\sqrt{2}}{5}$&  $ \frac{ \sqrt{2}}{5}$\\
$O^V$ & V&I&$\sqrt{2}$&0&0&$\sqrt{2}$\\
$O^W$ & III& II &0 & $\sqrt{\frac{2}{5}}$& -3$\sqrt{\frac{2}{5}}$&0\\
$O^X$ & IV & II & $\frac{2 \sqrt{2}}{5}$ & $\frac{\sqrt{2}}{5}$ &
$\frac{-3 \sqrt{2}}{5}$ & $\frac{-6 \sqrt{2}}{5}$ \\
$O^Y$ & VI & II & $\frac{ \sqrt{2}}{5}$ & $\frac{2\sqrt{2}}{5}$ &
$\frac{-6 \sqrt{2}}{5}$ & $\frac{-3 \sqrt{2}}{5}$ \\
$O^Z$ & V & II & $\sqrt{\frac{2}{5}}$ &0 & 0 &
 $-3\sqrt{\frac{2}{5}}$ \\
$O^a$& I & III & $\sqrt{2}$ &$\sqrt{2}$&0&0 \\
$O^b$& IV & III & $\frac{4}{\sqrt{5}}$ & $\frac{2}{\sqrt{5}}$&0&0 \\
$O^c$& VI & III & $\frac{2}{\sqrt{5}}$ & $\frac{4}{\sqrt{5}}$&0&0
\\
$O^d$ & I & IV & $\sqrt{\frac{2}{5}}$ & $\sqrt{\frac{2}{5}}$ &
$2\sqrt{\frac{2}{5}}$ & $2\sqrt{\frac{2}{5}}$ \\
$O^e$ & II & IV & $\sqrt{\frac{2}{5}}$ & -$\sqrt{\frac{2}{5}}$ &
-$2\sqrt{\frac{2}{5}}$ & $2\sqrt{\frac{2}{5}}$ \\
$O^f$ & VI & IV & $\frac{2}{5}$ & $\frac{4}{5}$ &$\frac{8}{5}$
&$\frac{4}{5}$ \\ 
$O^g$ &I & V& 0 & 0 & $\sqrt{2}$ & $\sqrt{2}$ \\
$O^h$ &II & V& 0 & 0 & $-\sqrt{2}$ & $\sqrt{2}$ \\
$O^i$ &III & V& 0 & 0 & 2 & 0\\
$O^j$ &IV & V& 0 & 0 & $\frac{2}{\sqrt{5}}$ & $\frac{4}{\sqrt{5}}$\\
\hline
\end{tabular}
\end{center}
\vspace{-0.3 cm}
\caption{{\small When the Higgs fields develop their VEVs, the
$n=1$ operators lead to the 
effective Yukawa couplings with
Clebsch coefficients as shown.}}
\label{tab:clebschn1}
\end{table} 
The $n=1$ operators are formed from different group theoretical
contractions of the indices in
\begin{equation}
O^{\alpha \rho y w}_{\beta \gamma x z} \equiv F^{\alpha a}
\bar{F}_{\beta 
x} h^y_a \bar{H}_{\gamma z} H^{\rho w}.
\label{n1ops}
\end{equation}
It is useful to define some SU(4) invariant tensors $C$, and SU(2)$_R$
invariant tensors $R$ as follows:
\begin{eqnarray}
(C_1)^{\alpha}_\beta &=& \delta^\alpha_\beta \nonumber \\
(C_{15})^{\alpha \rho}_{\beta \gamma} &=& \delta_\beta^\gamma
\delta^\rho_\alpha - \frac{1}{4} \delta_\beta^\alpha
\delta^\rho_\gamma \nonumber \\
(C_6)_{\alpha \beta}^{\rho \gamma} &=& \epsilon_{\alpha \beta \omega
\chi} \epsilon^{\rho \gamma \omega \chi} \nonumber \\
(C_{10})^{\alpha \beta}_{\rho \gamma} &=& \delta^\alpha_\rho
\delta^\beta_\gamma + \delta^\alpha_\gamma \delta^\beta_\rho \nonumber
\\
(R_1)^x_y &=& \delta^x_y \nonumber \\
(R_3)^{wx}_{yz} &=& \delta^x_y \delta^w_z - \frac{1}{2} \delta^x_z
\delta^w_y, \label{T1s}
\end{eqnarray}
where $\delta^\alpha_\beta$, $\epsilon_{\alpha \beta \omega \chi}$,
$\delta^x_y$, $\epsilon_{wz}$ are the usual invariant tensors of   
SU(4),
SU(2)$_R$. The SU(4) indices on $C_{1,6,10,15}$ are contracted with
the SU(4) indices on two fields to combine them into $\underline{1}$,
$\underline{6}$, $\underline{10}$, $\underline{15}$ representations of
SU(4) respectively. Similarly, the SU(2)$_R$ indices on $R_{1,3}$ are 
contracted with SU(2)$_R$ indices on two of the fields to combine them
into $\underline{1}$, $\underline{3}$ representation of SU(2)$_R$.

The SU(4) structures in Table~\ref{tab:clebschn1} are
\begin{eqnarray}
\mbox{I.} && (C_1)^\beta_\alpha (C_1)^\gamma_\rho \nonumber \\
\mbox{II.} && (C_{15})^{\beta \chi}_{\alpha \sigma} (C_{15})^{\gamma
\sigma}_{\rho \chi} \nonumber \\
\mbox{III.} && (C_6)^{\omega \chi}_{\alpha \rho} (C_6)^{\beta \gamma}_{\omega
\chi} \nonumber \\
\mbox{IV.} && (C_{10})^{\omega \chi}_{\alpha \rho} (C_{10})^{\beta
\gamma}_{\omega \chi} \nonumber \\
\mbox{V.} && (C_1)^\beta_\rho (C_1)^\gamma_\alpha \nonumber \\
\mbox{VI.} && (C_{15})^{\gamma \chi}_{\alpha \sigma} (C_{15})^{\beta
\sigma}_{\rho \chi}, \label{su4str}
\end{eqnarray}
and the SU(2) structures are
\begin{eqnarray}
\mbox{I.} && (R_1)^z_w (R_1)^x_y \nonumber \\
\mbox{II.} && (R_3)^{zq}_{wr} (R_3)^{xr}_{yq} \nonumber \\
\mbox{III.} && \epsilon^{xz} \epsilon_{yw} \nonumber \\
\mbox{IV.} && \epsilon_{ws} \epsilon^{xt} (R_3)^{sq}_{yr}
(R_3)^{zr}_{tq}  \nonumber \\
\mbox{V.} && (R_1)^z_y (R_1)^x_w  \nonumber \\
\mbox{VI.} && (R_3)^{zq}_{yr} (R_3)^{xr}_{wq}. \label{su2str}
\end{eqnarray}
The operators are
then given explicitly by contracting Eq.\ref{n1ops} with the
invariant tensors of Eq.\ref{T1s} given by Table~\ref{tab:clebschn1}
and Eqs.\ref{su4str},\ref{su2str}.
 

\newpage

\begin{center}
{\bf Appendix 2. Review of Analysis of Ref.\cite{422}}
\end{center}

In ref.\cite{422} we assumed 
that the Yukawa matrices at $M_{X}$ are all of the
form
\begin{equation}
\lambda^{U,D,E,N} = \left(
\begin{array}{ccc}
O(\epsilon^2) & O(\epsilon^2) & 0 \\ O(\epsilon^2) & O(\epsilon) &
O(\epsilon) \\ 0 & O(\epsilon) & O(1) \\
\end{array}\right),
\label{matrixform}
\end{equation}
where $\epsilon << 1$ and some of the elements may have approximate or
exact texture zeroes in them. 
First, we examine closer the assumption that the operator in the (33)
position of the Yukawa matrices is the renormalisable one. It has been
suggested in the past that the large value of $\tan \beta$ required by
the constraint
\begin{equation}
\lambda_t(M_{GUT}) = \lambda_b(M_{GUT}) = \lambda_\tau(M_{GUT})
\label{quadun}
\end{equation}
such as is predicted by the renormalisable operator, leads to some
phenomenological
problems. One such problem is that a moderate fine tuning mechanism is
required to radiatively break the electro-weak symmetry in order to
produce the necessary hierarchy of Higgs VEVs $v_1 / v_2\approx m_t /
m_b$~\cite{hitanbprobs},\cite{EWSB}. One could set about trying to
extend the present
model in a manner that would lead to an arbitrary choice of $\tan
\beta$, for example by introducing extra Higgs bidoublets. This route
has
its disadvantages in that a low value of $\tan \beta$ has been
shown~\cite{quad} in most schemes
to be inconsistent with $\lambda_b(M_{GUT}) = \lambda_\tau(M_{GUT})$
unification if the tau neutrino mass constitutes the hot dark matter
requiring
the Majorana mass of the right handed tau neutrino to be
$M^{\nu_\tau}_R \sim O(10^{12})$~GeV. To a very good
approximation, the largest diagonalised Yukawa coupling in $\lambda^I$
is equal to its 33 entry $\lambda^I_{33}$. 
(One may obtain small $\tan \beta$ solutions consistent
with $m_b$-$m_{\tau}$ unification and an intermediate neutrino
scale, in specific models: Either
large mixing in the
$\mu-\tau$ charged leptonic sector has to occur
\cite{geo2} or the Dirac-type Yukawa coupling of
the neutrino  has to be very suppressed \cite{dimp}.)

To force things to work
in a generic scheme, one solution could be to use a 
non-renormalisable operator in the 33
position which has some Clebsch factor $x>1$ such that
\begin{equation}
\lambda_t(M_{GUT}) = x \lambda_b (M_{GUT}) = x \lambda_\tau (M_{GUT}).
\label{3rdcl}
\end{equation}
Eq.\ref{3rdcl} would preserve the bottom-tau Yukawa unification, but
lower the prediction of $\tan \beta$ due to the bigger contribution to
the top Yukawa coupling. It may only be reasonable to examine $n=1$
operators in this context since we know that the third
family~\cite{422} Yukawa coupling is $\sim O(1)$ and higher
dimension operators could be expected to provide a big suppression
factor. Systematically examining the $n=1$ operators we find that only
the operator $O^{U}_{33}$, which leads to the prediction
\begin{equation}
\lambda_t (M_{GUT}) = 2 \lambda_b (M_{GUT}) = 2 \lambda_\tau(M_{GUT})
\end{equation}
can decrease $\tan \beta$. The change is minimal, from 56.35 to 55.19
for $\alpha_S(M_Z) = 0.117$ and $M^{\nu_\tau}_R = O(10^{12})$~GeV.
The reason that the change is minimal is due to the
fact that the Yukawa couplings are approximately at their quasi fixed
points \cite{topFP} and so even a large change to $\lambda_{t,b,\tau}
(M_X)$
produces only a small change in $\lambda_{t,b,\tau}(m_t)$, which are
the 
quantities that require a high $\tan \beta$ through the relations
in Eq.\ref{masspred}.
Another possibility would be to include $O_{33}^M,
O_{33}^V$ which would allow arbitrary $\tan \beta$ (in particular
intermediate $\tan \beta \sim 10-20$.) However, this would reduce the
predictivity of the scheme as $\tan \beta$ would become an input. One
might also be skeptical about whether a parameter $\sim 1$ could be
generated by a non-renormalisable operator in a perturbative scheme. It
would certainly require the heavy mass scales $M$ to be very close to
the VEVs $H,\bar{H},\theta,\bar\theta$ and we might therefore naively
expect large corrections to any calculation based on this model.
We thus abandon these ideas and continue with the usual renormalisable
operator in the 33 position of the Yukawa matrices that leads to
Eq.\ref{quadun}.  
We note in any case that a recent analysis~\cite{MGM} explains
that in gauge mediated supersymmetry breaking models, the radiative
mechanism
of electroweak symmetry breaking can be such that no fine tuning
occurs for large $\tan \beta$. In these models high $\tan\beta$
admits solutions of the hot dark matter problem in which the Yukawa
couplings unify \cite{quad}.

The hierarchy assumed in Eq.\ref{matrixform} allows us to   
consider the lower 2 by 2 block of the Yukawa matrices first.  In
diagonalising the lower 2 by 2 block separately, we introduce
corrections of order $\epsilon^2$ and so the procedure is consistent  
to first order in $\epsilon$.  We
found several maximally predictive ansatze that were constructed out
of the operators whose Clebsch coefficients are listed in 
table~\ref{tab:subset} for the $n=1$ operators. 
The explicit $n=1$
operators in component form are listed in the Appendix 1.
We label the successful lower 2 by 2 ansatze $A_i$:
\begin{eqnarray}
A_1 &=& \left[\begin{array}{cc} O^D_{22} - O^C_{22} & 0 \\ O^C_{32}
& O_{33} \end{array}\right] \label{A1} \\ A_2 &=&
\left[\begin{array}{cc} 0 & O^A_{23} - O^B_{23} \\ O^D_{32} & O_{33}
\end{array}\right]  \\
A_3 &=& \left[\begin{array}{cc} 0 & O^C_{23} - O^D_{23} \\ O^B_{32} &
O_{33}
\end{array}\right]  \\
A_4 &=& \left[\begin{array}{cc} 0 & O^{C}_{23} \\ O^A_{32} - O^B_{32}
& O_{33}
\end{array}\right]  \\
A_5&=& \left[\begin{array}{cc} 0 & O^{A}_{23} \\ O^C_{32} - O^D_{32} &
O_{33}
\end{array}\right] \\
A_6&=& \left[\begin{array}{cc} O^{K}_{22} & O^{C}_{23} \\ O^M_{32} &
O_{33}
\end{array}\right]\\
A_7&=& \left[\begin{array}{cc} O^{K}_{22} & O^G_{23} \\ O^G_{32} &
O_{33}
\end{array}\right]  \\
A_8&=& \left[\begin{array}{cc} 0 & O^H_{23} \\ O^G_{32} - O^{K}_{32} &
O_{33}
\end{array}\right].
\label{endans}
\end{eqnarray}
We now note that solutions $A_{2-8}$ require a parameter $H_{23} \sim
O(1)$ to attain the correct $\lambda_\mu$ and $V_{cb}$. Any
calculation based on the hierarchy assumed in Eq.\ref{matrixform} is
therefore inconsistent and so we discard these solutions.
We also note that $O_{32}$ only has the effect of fixing $V_{cb}$ to a
good approximation and so can consist of any operator in
Table~\ref{tab:clebschn1} that has a different Clebsch coefficient for
up quark and down quark Yukawa couplings. The precise operator
responsible for $V_{cb}$ has no bearing on the rest of the calculation
and we therefore just make an arbitrary choice of $O^C_{32}$ for the
rest of this paper. We also note that for the phenomenologically
desirable and predictive relation
\begin{equation}
\frac{\lambda^D_{22}(M_{GUT})}{\lambda^E_{22}(M_{GUT})} = 3,
\label{lspred}
\end{equation}
to hold, we may replace $O_{22}^D-O^C_{22}$ in $A_1$ with
$O^W_{22}+O^C_{22}$,
$O^X_{22}+O^D_{22}$ or any other combination of two operators which
preserves Eq.\ref{lspred} and allows $\lambda_{22}^U$ to be smaller
and independent of $\lambda^{D,E}_{22}$. In fact, the preferred
solution is that the dominant operator in that position be $O_{22}^W$
which does not give a contribution to the up quark mass. Then, a
subdominant operator would be responsible for the entry
$\lambda_{22}^U$ and would therefore be suppressed naturally by one or
more powers of $\epsilon$. 

\begin{center}
{\bf Appendix 3. Flatness Conditions in the String Model.}
\end{center}

 We give here the constraints  on the various singlet  VEVs 
obtained from the F and D flatness
conditions in the string
spectrum of the model in section 8.
{}From the F-flatness of the superpotential
one derives 18 conditions, 
which  are the following :
\begin{eqnarray}
\bar\xi_1\bar\xi_4 & = & 0 \nonumber\\
\xi_1\xi_4 & = & 0 \nonumber \\
\xi_2\bar\xi_3+\zeta_1^2+\zeta_2^2 & = & 0 \\
\bar\xi_2\xi_3+\bar\zeta_1^2+\bar\zeta_2^2 & = & 0
\nonumber\\
\xi_i\bar\xi_i+\zeta_1 \bar{\zeta}_1
+\zeta_2 \bar{\zeta}_2 & = & 0 \nonumber \\
\zeta_1\bar\zeta_2+\bar\zeta_1\zeta_2 & = & 0 \nonumber \\
2 \bar{\Phi}_{\overline{12}} \zeta_1 +
\frac{1}{2} \Phi_3 \bar{\zeta}_1 + \Phi_4 \bar{\zeta}_2
 & = & 0
\nonumber \\
2 {\Phi}_{\overline{12}} \bar \zeta_1 +
\frac{1}{2} \Phi_3 {\zeta}_1 + \Phi_4 {\zeta}_2
 & = & 0
\nonumber \\
2 \bar{\Phi}_{\overline{12}} \zeta_2 +
\frac{1}{2} \Phi_3 \bar{\zeta}_2 + \Phi_4 \bar{\zeta}_1
 & = & 0
\nonumber \\
2 {\Phi}_{\overline{12}} \bar \zeta_2 +
\frac{1}{2} \Phi_3 {\zeta}_2 + \Phi_4 {\zeta}_1
 & = & 0
\nonumber \\
\bar{\Phi}_{12} \xi_4 + \frac{1}{2} \Phi_3 \bar{\xi}_1
 & = & 0
\nonumber \\
{\Phi}_{{12}} \bar{\xi}_4 + \frac{1}{2} \Phi_3 \xi_1
 & = & 0
\nonumber \\
\bar{\Phi}_{\overline{12}} \bar{\xi}_3 + \frac{1}{2} \Phi_3 \bar{\xi}_2
  & = & 0 \nonumber \\
{\Phi}_{\overline{12}} {\xi}_3 + \frac{1}{2} \Phi_3 {\xi}_2
 & = & 0 \nonumber \\
{\Phi}_{\overline{12}} \bar{\xi}_2 + \frac{1}{2} \Phi_3 \bar{\xi}_3
 & = & 0 \nonumber \\
\bar{\Phi}_{\overline{12}} {\xi}_2 + \frac{1}{2} \Phi_3 {\xi}_3
 & = & 0 \nonumber \\
\bar{\Phi}_{12} \xi_1 + \frac{1}{2} \Phi_3 \bar{\xi}_4
 & = & 0
\nonumber \\
{\Phi}_{{12}} \bar{\xi}_1 + \frac{1}{2} \Phi_3 \xi_4
 & = & 0\label{flcon}
\end{eqnarray}
Now, a possible choice of non- zero singlet VEVs which satisfy
the system (\ref{flcon}), is 
\beq
<\Phi_{12}>\; , <\bar\Phi_{12}^->\; , <\xi_1>\; , <\bar\xi_2>\; \not= 0
\label{schoice}
\eeq
accompanied with non - zero VEVs  of the following two
hidden (octets under $SU(8)_h$) fields 
\beq
<Z_5>\; , <\bar{Z}_3'>\; \not= 0
\label{hchoice}
\eeq
Taking all other singlet and hidden field VEVs equal to zero,
the D- flatness conditions  read \cite{lt}
\begin{eqnarray}
|Z_5|^2-2|\bar\Phi_{12}^-|^2-|\xi_1|^2-|\bar\xi_2|^2
+\frac{3\alpha_u}{2\pi}   & = &  0\label{anom1}\\
\frac 12|Z_5|^2-|\bar\xi_2|^2   & = &   0\label{nona2}\\
2|\xi_1|^2-|\bar\xi_2|^2-2|\bar\Phi^-_{12}|^2
+\frac 32|\bar{Z}_3'|^2 +|Z_5|^2 &   = &   0\label{nona3}\\
2|\Phi_{12}|^2+|\xi_1|^2-\frac 12|\bar{Z}_3'|^2  & = &   0
\label{nona4}
\end{eqnarray}
The scale of the non- zero singlet VEVs is determined by the above 
conditions. There are five equations to determine four parameters, 
thus one has the freedom to fix one of the non zero VEVs in 
Eqs.(\ref{schoice},\ref{hchoice}) from phenomenological requirements.
In any case, from the above equations it turns out that
the natural scale of the non zero VEVs are
of the order $\frac{\alpha_u}{\pi}M_{Pl}$. For $\alpha_u\sim 10^{-1}$
one can see that their magnitude  is of the required order
 to contribute in the mass operators.

\end{document}